\documentclass[12pt,preprint]{aastex}

\newcommand{\etal}{{\em et al.}}            

\newcommand{\xmm}{{XMM-Newton}}
\newcommand{\chdr}{{\em Chandra}}

\shorttitle{X-rays from Cen A ISM and Radio Lobes}
\shortauthors{Kraft \etal}

\begin{document}

\title{X-ray Emission from the Hot ISM and SW Radio Lobe of the Nearby Radio Galaxy Centaurus A}
\author{R. P. Kraft, S. V\'{a}zquez\altaffilmark{1}, W. R. Forman, C. Jones, S. S. Murray}
\affil{Harvard/Smithsonian Center for Astrophysics, 60 Garden St., MS-31, Cambridge, MA 02138}
\author{M. J. Hardcastle, D. M. Worrall}
\affil{University of Bristol, Department of Physics, Tyndall Ave., Bristol BS8 ITL, UK}
\author{E. Churazov}
\affil{Max Planck Institute f\"{u}r Astrophysik, 85740 Garching bei M\"{u}nchen, Germany and Space Research Institute (IKI), Moscow 117810, Russia}

\altaffiltext{1}{Present address: Department of Physics, Brioda Hall, University of California, Santa Barbara 93106}

\begin{abstract}
We present results from two {\em Chandra}/ACIS-I observations
and one {\em XMM-Newton} observation of X-ray emission from the ISM and the inner radio lobes of
the nearby radio galaxy Centaurus A.  The ISM has an average radial surface brightness profile that is
well described by a $\beta$-model profile with index $\beta$=0.40$\pm$0.04
and a temperature of $k_BT_{ISM}\sim$0.29 keV beyond 2 kpc from the nucleus.
We find that diffuse X-ray emission is coincident with
the outer half of the southwest radio lobe, and a bright X-ray enhancement is detected
along the edge of the lobe.  On the basis of energetic and lifetime arguments, we
reject a nonthermal explanation for this emission.
We model this emission as a thin, hot shell or cap of X-ray
emitting plasma surrounding the radio lobe that was created by the
supersonic inflation of the lobe.
This plasma shell is both hotter than ($k_BT_{SH}\sim$2.9 keV)
and greatly overpressurized relative to the ambient ISM
indicating supersonic expansion.  We estimate that the lobe is expanding into the ISM at
approximately Mach 8.5 or 2400 km s$^{-1}$.  We are not directly observing the bow shock,
but rather the cooler, denser material that is accumulating ahead of the contact discontinuity.
The thermal energy in the shell is a significant fraction of the thermal energy
of the hot ISM, demonstrating the possibility that the hot ISM of early
galaxies can be re-energized by outflows from nuclear activity.
Interestingly, no similarly bright X-ray emission is detected in or
along the edge of the NE lobe, implying that there are differences in the dynamics and evolution
of the kpc-scale radio components.  
\end{abstract}

\keywords{galaxies: active - galaxies: individual (Centaurus A, NGC 5128) -
galaxies: jets - X-rays: galaxies - galaxies: ISM - radio continuum: galaxies}

\section{Introduction}

Centaurus A (NGC 5128, Cen A) is the nearest (d$\sim$3.4 Mpc, $1'\sim$1 kpc) active galaxy to
the Milky Way, and because of its proximity has been well studied across the
electromagnetic spectrum \citep{isr98}.  Optically, Cen A is an elliptical galaxy crossed by a
dust lane, thought to be the result of a merger with a small spiral galaxy \citep{sch94}.
Radio observations of Cen A show a bright nucleus,
a milliarcsecond-scale jet and counter jet, a one-sided
kiloparsec scale jet NE of the nucleus, two radio lobes
(the inner radio lobes) NE and SW of the nucleus, and
extended, diffuse emission spanning an 8$^\circ\times$4$^\circ$
region on the sky \citep{bur83,cla92,tin98,isr98}.
Earlier X-ray observations of Cen A found a complex
morphology with emission from several distinct components including
the active nucleus, the jet, the hot ISM, and a population of
X-ray binaries \citep{fei81, dob96}.  ROSAT HRI and EXOSAT LE observations
detected X-ray emission from the vicinity of the southwest radio
lobe \citep{dob96,mor89}, but the limited spectral resolution
of these observations and possible confusion with a bright foreground star
made interpretation uncertain.

The {\em Einstein} Observatory first detected extended X-ray emission from
elliptical and S0 galaxies, and since then the spectra and morphology
of these objects has been extensively studied with ROSAT and ASCA
\citep{for85,tri85,tri91,mat97,fin00}.  For the most massive, luminous, early-type galaxies, this
emission originates in a hot, low-density corona that is bound by the
gravitating dark-matter potential.  The X-ray spectra of these
halos are generally well described by a single-temperature optically thin
plasma model with depleted abundance values with respect to
solar of Z$\sim$0.5.  Typical temperatures and central densities
of this gas are $\sim$1 keV and 0.1 cm$^{-3}$, respectively \citep{mat97}.  The cooling
time of the gas in the central regions of these galaxies is $\sim$10$^8$ yrs, implying that
in the absence of any energy input, the gas will cool and condense onto the central black hole.
For less massive early-type galaxies, such as Cen A, the situation is
somewhat more complex as the contribution of
a population of LMXBs becomes increasingly more important to the 
overall X-ray emission \citep{fab89,bro01,sar01}.
{\em Chandra} and {\em XMM-Newton} have greatly increased our understanding
of these objects through spatially resolved measurements of the
temperatures and elemental abundances from which the characteristics of
dark matter halos can be deduced \citep{loe02}, and by resolving out the 
LMXB population in nearby objects \citep{bla01,sar01}.

Recent {\em Chandra} observations of the environments of several radio galaxies,
notably NGC 4636 \citep{jon02}, M84 \citep{fin01}, Hydra A \citep{mcn00,dav01}, Perseus A
\citep{fab00}, 3C 317 \citep{bla01r}, and M87 \citep{chu01}, have
demonstrated a complex interaction between the relativistic
plasma of the radio lobes and the thermal, X-ray 
emitting gas of the interstellar or intracluster medium.
There has also been considerable theoretical interest in
this subject \citep{hei98,brg02m,brg02n,rey01}.
In several cases, the thermal gas appears to have been
displaced by the expansion of the radio lobes,
creating a cavity or hole in the X-ray emission
from the interstellar or intracluster medium.  In some cases,
X-ray emitting shells have been observed surrounding this
cavity.  In most cases where such shells have been detected, they
are cooler than the surrounding medium and are thought to
the result of entrainment of cool material from the central regions of the galaxy or
cluster by the inflation of the lobe.
In a few cases (NGC 4636 and Cyg A), there is some
evidence that the shells are hotter than the ambient
medium and have been shock heated by the supersonic
inflation of the radio lobe.

One important consequence of this interaction, particularly
if the expansion is supersonic, is that
the kinetic energy of the particles in the radio lobes could be
transferred to the thermal medium thereby heating it.  The energy
in shock-heated gas would be transferred to the ISM via conduction, perhaps
even in the presence of magnetic fields
\citep{nar01}.  This process could
be particularly important in understanding the dynamics
of the central regions of cooling-flow clusters, as
the thermal lifetime of the centrally condensed material 
present in these objects is less than the Hubble time and heat
conduction is less effective in the cores.
The gas must therefore be occasionally reenergized during its
lifetime.  Radio galaxies are one possible source for this
reheating.  Unfortunately, there has been little direct evidence for
heating of the interstellar or intracluster medium by the radio plasma 
for any of the sources listed above, and the relationship between
nuclear outflows, nuclear activity, and cooling flows is uncertain.
It has been suggested that there is a cyclical relationship between
the cooling of the X-ray emitting corona and galaxy activity \citep{tuc97,cio97,cio01}.
Clearly the relationship between the hot ISM, radiation from
the central black hole, and outflows of relativistic plasma
generated by nuclear activity is very complex.

In this paper, we present the results from two {\em Chandra}/ACIS-I
observations and an {\em XMM-Newton} observation of
X-ray emission from the ISM and from the radio lobes of Cen A.  
The primary goals of this work are to determine the thermodynamic state of the hot
ISM in Cen A and to better understand the energetics and dynamics of the
interaction between the relativistic plasma of the radio lobes and the X-ray emitting corona.
In this paper, we present the temperature and surface
brightness profile of the ISM, and discuss the nature of the X-ray enhancement
associated with the southwest radio lobe.  
We demonstrate that it is unlikely that this enhancement is non-thermal in nature,
but most likely originates from compression and shock-heating
of the hot ISM via the supersonic expansion/inflation of the radio lobe.
This is the fifth paper in our series on {\em Chandra} and
{\em XMM-Newton} observations of Cen A.
In four previous papers on Cen A, we presented results of
a {\em Chandra}/HRC observation \citep{kra00},
and results of the two deeper {\em Chandra}/ACIS-I observations
on the X-ray point source population \citep{kra01},
the X-ray jet \citep{kra02}, and complex morphology of the ISM \citep{kar02}.
Future publications will present an analysis and
discussion of the spectrum of the active nucleus (Kraft \etal, in preparation),
a detailed comparison of high-resolution X-ray and radio observations
of the jet (Hardcastle \etal, in preparation), and a more detailed analysis
of the morphology and dynamics of the hot ISM (Karovska \etal, in preparation).
This paper is organized as follows.
Section 2 contains a brief summary of the observations
and the instrumentation.  Spectra and images of the ISM and the radio lobes are
presented and discussed in sections 3 and 4, respectively.
We end with a summary and conclusions in section 5.
We use J2000 coordinates throughout the paper.

\section{Instrumentation and Observations}

Cen A was observed twice with the {\em Chandra}/ACIS-I instrument,
on December 5, 1999 and May 17, 2000, and once with {\em XMM-Newton}
on February 2, 2001.  The observation times were 35856 $s$ and 36510 $s$
for the two {\em Chandra}, and 23060 $s$ with {\em XMM-Newton} with the
medium optical blocking filter inserted.  A summary
of the observational parameters is contained in Table~\ref{obslog}.
Descriptions of the instrumental capabilities of the
two observatories are presented elsewhere \citep{wei00,pog03,jan01}.
The absolute position on the sky of the {\em Chandra}
observations was determined by comparison of the X-ray point
sources at the edge of the FOV with stars in the USNO catalog
(see \citet{kra01} for details).  Then the absolute
position of the {\em XMM-Newton} observation was determined
by comparing the positions of X-ray point sources with {\em Chandra}
positions.  A considerable ($\sim 12''$) correction was applied to
the {\em XMM-Newton} set as a result.
After correction, both data sets are aligned on the sky
and with respect to each other to better than $1''$.
We have generally relied on the {\em XMM-Newton} data
for spectral analysis of extended low surface brightness
features, but we have used the {\em Chandra}/ACIS-I data to
determine the positions of point sources.

The {\em Chandra} raw event table was
filtered to include only grades 0,2,3,4, and
6.  All events below 0.4 keV and above 5 keV were removed.  The response
of the ACIS-I drops rapidly below 0.4 keV, so that most of the
events below this are background.  Above 5 keV,
most of the events are either particle background events or
events in the PSF wings from the bright nucleus.
All events at node boundaries were removed because of uncertainties
in grade reconstruction.  All events with pulse-height invariant
channel (PI) equal to 0, 1, or 1024 were removed as they represent
unphysical signals; hot CCD columns and pixels were removed.  Short-term
transients due to cosmic rays which produced events in three or more consecutive frames
that could mimic a point source (van Speybroeck 2000, private
communication) also were removed.
Data from the two observations were co-added to create images,
but not for spectral analysis for two reasons.  First, the focal-plane
temperatures differed (-110$^\circ$C during the first observation, and
-120$^\circ$C during the second).  Second, differences in roll angles
of the satellite and pointing directions between the observations
resulted in the southwest lobe being placed along the edge of the I3
chip for the first observation and on the I0 chip close to the best
focus of the telescope for the second.
All images presented below were exposure corrected. 

Data from all three CCD imaging spectroscopy instruments (EPIC/MOS1, MOS2, and the
PN camera) from the {\em XMM-Newton} observation were used in this analysis.
The EPIC/MOS events tables were filtered to include only events
with patterns 0 through 12, and with the FLAG parameter less
than or equal to 1.
The EPIC/PN events tables were filtered to include only events
with pattern 0, and with the FLAG parameter less than or equal to 1.
All events with PI value greater than 12000 (i.e. energy greater
than 12 keV) were also removed as the telescope has no
response to X-rays above this energy.
The data from MOS1 and MOS2, but not from the PN camera, were combined to
create images as shown below.  All spectral fits were done on the three data sets independently.
Response matrices and ancillary response files
were generated using the {\em arfgen} and
{\em rmfgen} tools in the SAS software ($xmmsas\_20010917\_1110-5.2.0$).
All spectral fitting (both {\em Chandra} and {\em XMM-Newton}) was performed using 
the XSPEC (V 10.0) software package.  
All exposure corrections made to the {\em XMM-Newton}
were done using the exposure maps provided with the pipeline processed data.

\begin{figure}
\plotone{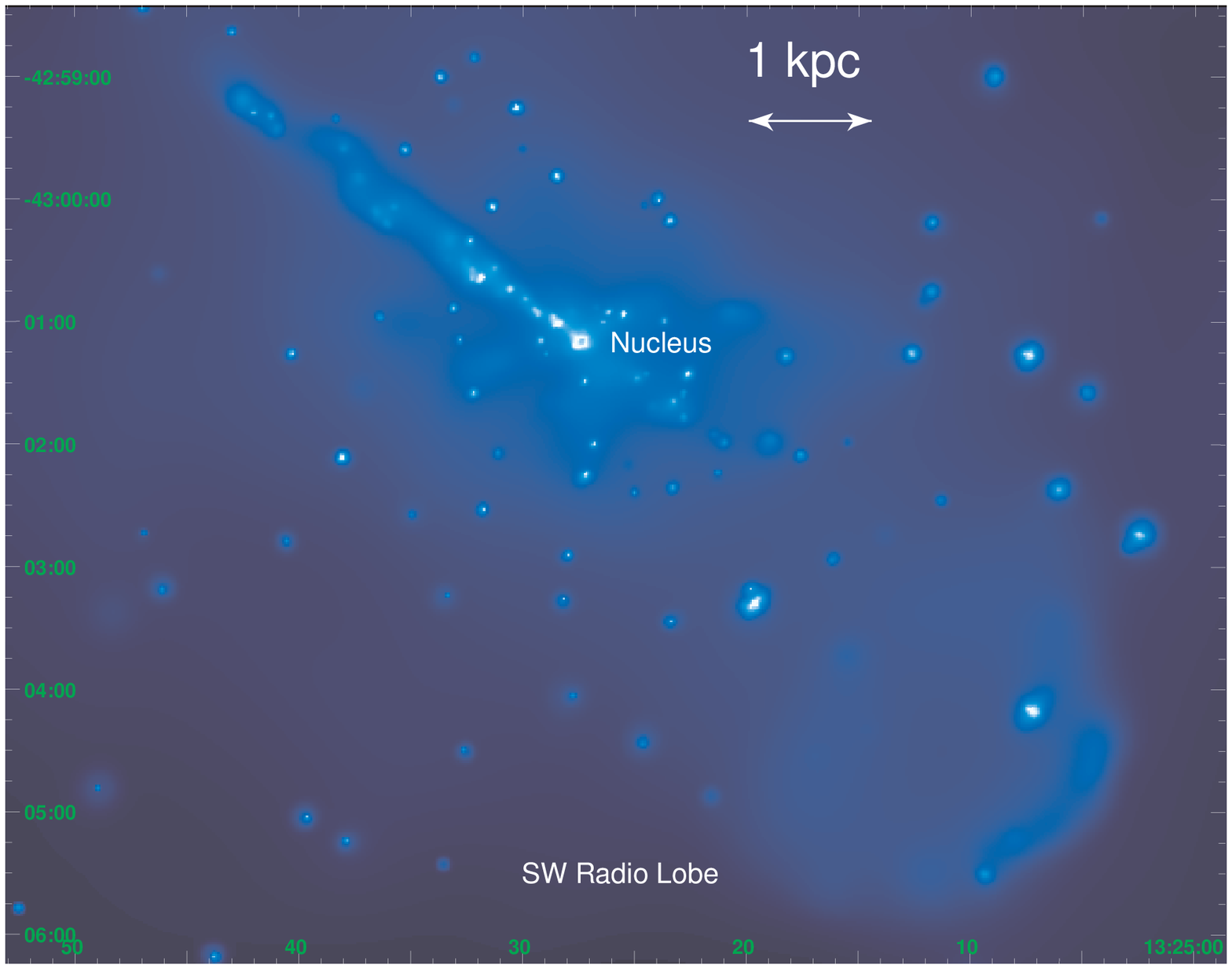}
\caption{Adaptively smoothed, co-added, exposure-corrected {\em Chandra}/ACIS-I
X-ray image of Centaurus A in the 0.4-2 keV band.  The nucleus is
the bright source near the center of the field, and the jet
extends to the NE.  North is up and east is left.}\label{chdimg}
\end{figure}

Co-added, exposure corrected, adaptively smoothed X-ray images
of Cen A from the {\em Chandra} and from the PN camera/{\em XMM-Newton} observations
are shown in Figures~\ref{chdimg} and~\ref{xmmrad}, respectively.
Radio contours (13 cm - Gaussian width of beam 
$30.45''$(RA) $\times$ $20.31''$(DEC) taken with the Australia Telescope Compact Array (ATCA))
have been overlaid onto Figure~\ref{xmmrad}.
The nucleus, the jet NE of the nucleus and the two
inner lobes of Cen A are clearly visible in the radio contours.
The bright active nucleus, the jet, the diffuse emission from the
ISM, and many point sources (mostly XRBs within Cen A) are clearly
visible in the X-ray images.  There is also an X-ray enhancement along the
edge of the southwest radio lobe, and an excess of diffuse emission (above that
of the ISM) in the interior of the lobe.
The close alignment of the X-ray enhancement just at the edge
of the lobe with the lowest radio contours strongly argues that they are related.
No significant excess or deficit of emission (i.e. above or below that
expected from the hot ISM) is detected in the vicinity of the NE radio lobe.

\section{The Interstellar Medium}

\subsection{Data Analysis}

To determine the temperature and radial surface-brightness profile of the
coronal X-ray emission as a function of distance from the nucleus,
it is necessary to exclude all of the emission from the point
sources and to carefully estimate the background.  The bulk of the
emission from the ISM of Cen A is below 1 keV where {\em XMM-Newton} has considerably
more effective area than {\em Chandra}/ACIS-I. 
For this and another reason described below, we rely primarily 
on the {\em XMM-Newton} data for determination of the parameters of the ISM.
We used the {\em Chandra} data to determine the positions of the
point sources, and then excluded a region of $25''$ radius around these positions
in the {\em XMM-Newton} data.  The jet and southwest radio lobe also were excluded
from all analysis of the ISM.

\begin{figure}
\plotone{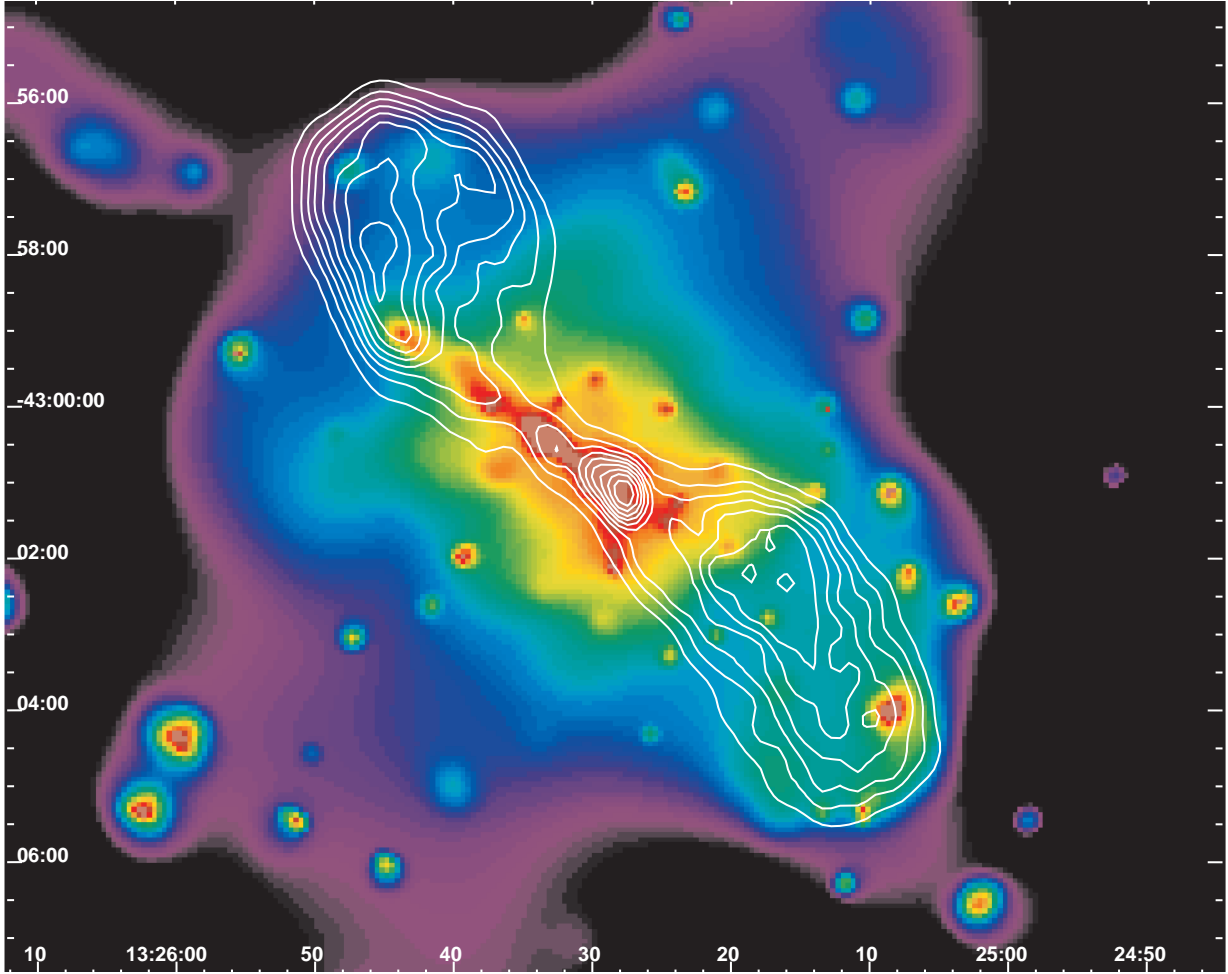}
\caption{Adaptively smoothed, co-added (MOS1 and MOS2), exposure-corrected
{\em XMM-Newton} X-ray image
of Centaurus A in the 0.5-2 keV band with 13 cm radio contours
overlaid.}\label{xmmrad}
\end{figure}

Careful background subtraction is critical to the accurate determination of
spectral parameters and to the radial surface-brightness profile.  
In the central regions of the galaxy, the emission of the ISM is seen clearly above the
background (see Figures~\ref{chdimg} and~\ref{xmmrad}).
In our initial analysis, we attempted to use the standard background files that are available online
for both {\em Chandra} and {\em XMM-Newton}.  These turned out to be entirely inappropriate
for two reasons.  First, they are generated from several high galactic latitude observations and
Cen A is located in the North Polar Spur (NPS) (l=309.52, b=+19.42).  The intensity and spectrum of the XRB
of the NPS region is considerably different than that at high galactic latitude \citep{gar92,sno97}.
Second, the standard background files for {\em XMM-Newton} were created from observations using the 
thin optical blocking filter, and our observation was made with the medium optical
blocking filter to avoid contamination from Cen A.

It is therefore necessary to use a locally determined background at the edge of the FOV.
Unfortunately Cen A is so close to us that diffuse X-ray emission fills the entire FOV of
both observatories.  At the edge of the FOV, the surface brightness of the ISM is still
a significant fraction ($\sim30$\%) of the total (background+source) surface brightness.
Fortunately, the background of both observatories below 1 keV is dominated by emission from the
diffuse X-ray background (XRB), not charged particles, so that we could model the
spatial variation of the background by the telescope vignetting function.
The measured spatial variation of the background at low energies is somewhat
flatter than what one would expect based on the vignetting.  We have used
the results of analysis of the low energy background
in the PN camera to model this effect
\footnote{For further details on the {\em XMM-Newton} background,
see $http://wave.xray.mpe.mpg.de/xmm/cookbook/$}.

Any region along the edge of the FOV contains both source and background, so it is therefore
necessary to iteratively estimate the contributions from source and
background at the edge of the FOV.  We selected regions in the PN and MOS
data $\sim$14$'$ NW of the nucleus
and fit a four component model with two MEKAL plasmas plus a power-law
with Galactic absorption (N$_H$=7$\times$10$^{20}$ cm$^{-2}$) and a
neutral Al K$_\alpha$ line at 1.49 keV (an instrumental artifact),
to model the background.  The best-fit temperatures for the thermal
components are consistent with previous measurements
of the NPS \citep{sno97,caw98}.  This model is an overestimate of the true background
because it contains some flux from Cen A as well.
Using this background, we then determined
the temperature of the corona as a function of distance in five annular bins
each $2'$ wide centered on the nucleus.  The parameterized background model, appropriately
scaled for solid angle, was included in the spectral fitting as a fixed component.
For the coronal emission from Cen A, we assumed a MEKAL model with depleted (Z=0.4) abundances
and Galactic absorption.  We performed the fits with the abundance as a free parameter as well,
but at the temperature of the ISM ($\sim$ 0.3 keV), the emission is line dominated and the
continuum level poorly constrained.
The abundance can be traded against the normalization and is thus poorly constrained.
Except in the central $2'$ bin where variable absorption from the dust lane hardens
the spectrum, all fits indicate a temperature between 0.27 and 0.32 keV, with a small
decrease as a function of distance from the nucleus.
We note that the temperature probably does rise toward the center, but this is
difficult to quantify because of the complex morphology due to
the variable absorption of the dust lane.

We then computed the radial surface brightness profile of the coronal emission in the 0.4 to 1.0 keV band
excluding all point sources, the jet, the southwest radio lobe, and subtracted the
(overestimated) background described above.
We fit a $\beta$-model of the form
\begin{equation}
S(r)=S_0\left(1+(\frac{r}{a})^2\right)^{-3\beta +{\frac{1}{2}}}, \label{sbeq}
\end{equation}
to the profile where $S_0$ is the surface brightness at the center of the galaxy, $r$ is the distance
from the center of the galaxy, and $a$ and $\beta$ are parameters determined by
fitting.  The electron density is then given by
\begin{equation}
n(r)=n_0\left(1+(\frac{r}{a})^2\right)^{-3\beta/2 }, \label{denseq}
\end{equation}
where $n_0$ is the central density.
The central density can be computed from the surface brightness
profile and the integrated luminosity.
The luminosity within a radius $R$ is then given by \citep{dav90}
\begin{equation}
L(R)={\frac{2\pi n_e n_H \Lambda_0 a^3}{(1-3\beta)}}\int_{0}^{\infty} ((1+s^2+(\frac{R}{a})^2)^{-3\beta +1} - (1+s^2)^{-3\beta +1})ds, \label{lumeq}
\end{equation}
where $\Lambda_0$ is the radiative cooling coefficient (ergs cm$^3$ s$^{-1}$),
$n_e$ and $n_H$ are the central electron and proton densities, respectively,
and $s$ is the distance along the line of sight in units of $a$.
The parameter $\beta$ is determined by a least squares fit to the
radial surface brightness profile, and the normalization of the
density profile, $n_0$, is determined by converting the observed X-ray flux
to luminosity using XSPEC and solving the integral in Equation~\ref{lumeq}.
Throughout this paper, we have assumed that $n_e \sim 1.18 n_H$, which is
appropriate for the sub-solar abundance plasma.

Because of the complex structure and variable absorption by the dust lane in the central $2'$
of the galaxy, the parameter $a$, the core radius from
Equation 1, is not constrained.  There is
evidence that the temperature of the gas in the central $2'$ of the
galaxy is somewhat hotter than the outer regions, so the isothermal
approximation is not appropriate.  Even detailed deprojection analysis might
not fully account for the complex three dimensional structure of the X-ray
emission.  We therefore decided to fix $a$ at 0.5 kpc ($0.5'$) and fit the profile
for $\beta$ between $2'$ and $10'$ from the nucleus.
The parameters $a$ and $\beta$ are generally strongly coupled \citep{dav01}, but we are fitting
the surface brightness profile in the region where it is well modeled by
a power-law (i.e. $r >> a$), so that $\beta$ is well constrained, and
so is $n_H(r)$ far from the nucleus.  This last point is of particular
importance in the discussion of the southwest radio lobe below.

\begin{figure}
\plotone{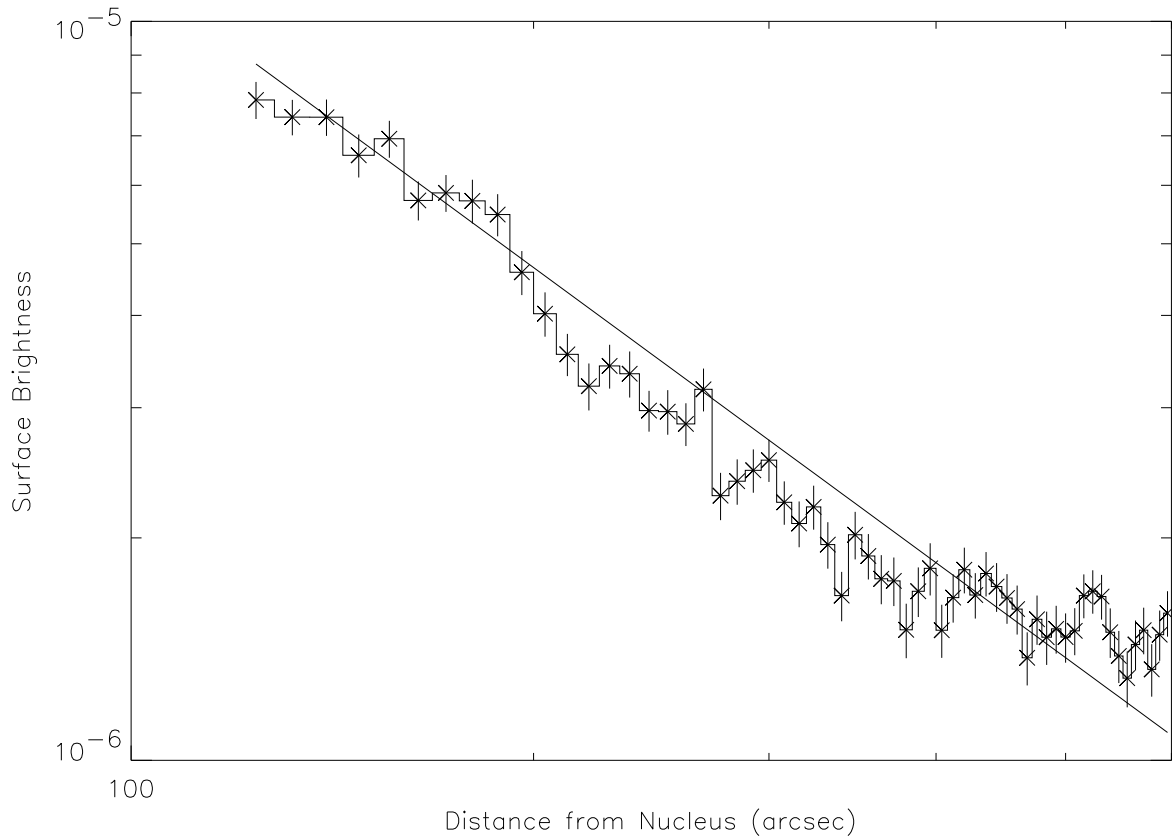}
\caption{Radial surface brightness profile in the 0.4 to 1.0 keV
band (histogram) and best-fit $\beta$-model (smooth curve)
of the PN camera data of the ISM of Cen A with 1$\sigma$ counting
statistics error bars.  The reduced $\chi^2$ of the $\beta$-model fit
is 2.1, indicating a marginal fit.  This is because of
the azimuthal dependence of both the surface brightness and temperature.}\label{sbp}
\end{figure}

\begin{figure}
\plotone{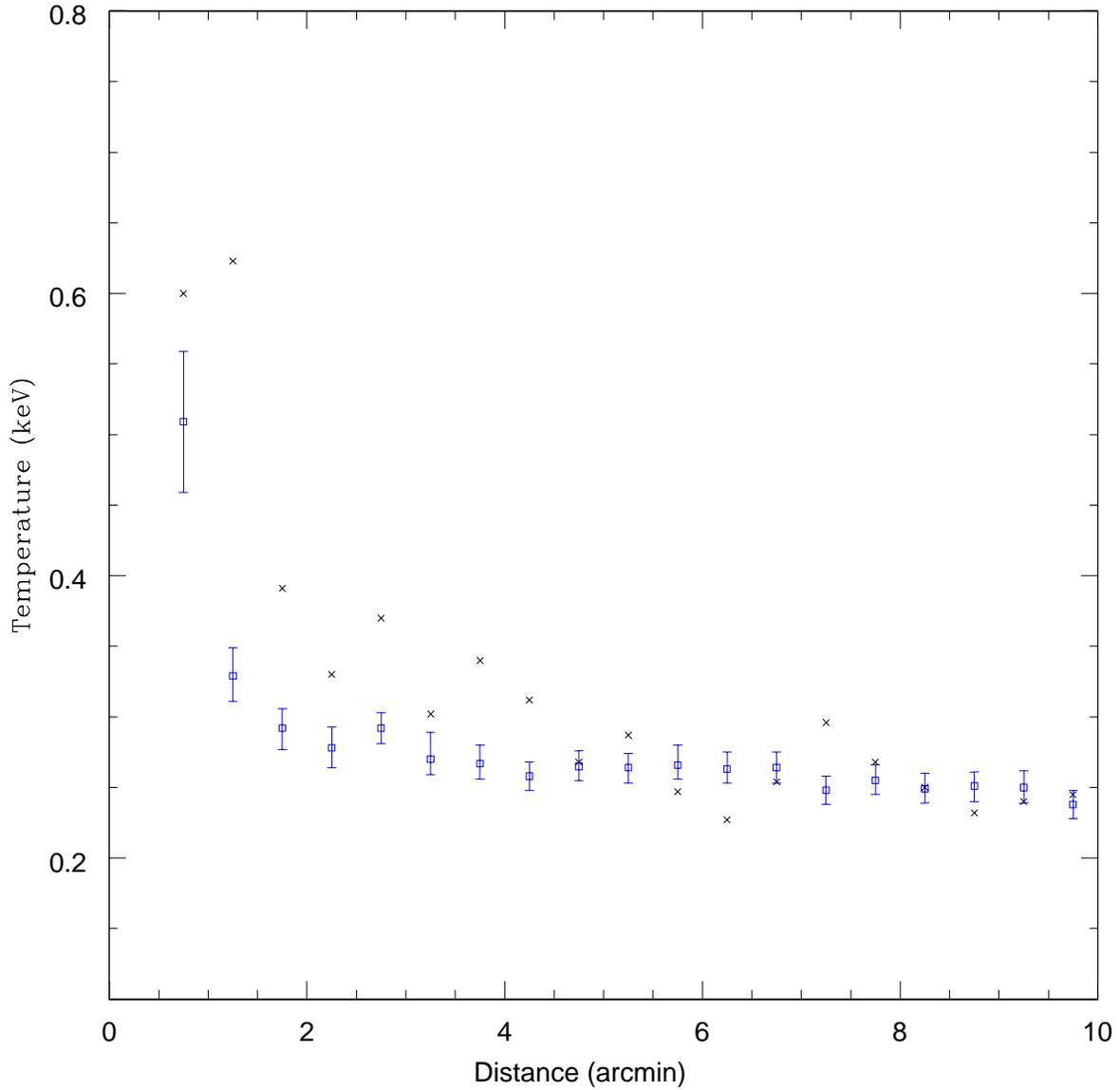}
\caption{Best fit single-temperature profile of hot ISM of Cen A
as a function of distance from the nucleus.  The blue points with the
errorbars and the black points without the errorbars are the
the best fit values from the PN and MOS cameras, respectively.  The
errors on the MOS points are comparable to the PN data but are not
shown for clairity.}\label{tprof}
\end{figure}

The best-fit $\beta$ model was extrapolated to the position of the background region
to determine the contribution of emission from Cen A in the background 
estimate.  Approximately 30\% of the flux in this region originates from
the galaxy and is not true background.  Using an appropriately reduced background,
the radial surface brightness profile was recomputed.  The radial surface brightness profile and the temperature
profile using the iteratively determined background
are shown in Figures~\ref{sbp} (PN camera only) and~\ref{tprof} (PN and
MOS cameras, respectively.
Implicit in the interpretation of the radial profile is that the temperature of the ISM in
the background region is not significantly different than that
closer to the galaxy.  Beyond the central region of the galaxy, the temperature of the ISM is
slowly decreasing with increasing radius, but we will assume a constant temperature
of $\sim$0.29 keV in the analysis below.  

The best-fit values of $\beta$ are 0.39$\pm$0.04 and 0.40$\pm$0.04 for the PN and MOS images,
respectively.  The fit was made between $2'$ and $10'$ from the nucleus
to avoid complications due to variable absorption by the dust lane and
the possible temperature rise in the central regions of the galaxy.
The uncertainty in $\beta$ is dominated by
uncertainty in the estimate of the background.
The reduced $\chi^2$ of the fit is 2.1 and 1.8 for the PN and MOS cameras, respectively, indicating
a marginal fit.  This is due to the azimuthal structure in both the
surface brightness (see Figure~\ref{xmmrad}) and temperature (see discussion
below).  For the purposes of understanding the dynamics of the radio
lobe discussed below, these variations are fairly
large (up to a factor of 2 in surface brightness or 40\% in particle density)
but we will use the azimuthally averaged $\beta$-model for estimates
of the density and pressure of the ISM.
The central density of the ISM for the assumed value of
$a$=0.5 kpc is $n_H(0)$=(3.7$\pm 0.4)\times$10$^{-2}$ cm $^{-3}$.

We have performed a similar analysis on the {\em Chandra}
data, but derive a higher temperature of about $\sim$0.4-0.45 keV.
We determined the background in a similar manner to that described
above using data from
the S2 chip, and other than normalization, we derive similar spectral
parameters from this background data set as we did from the {\em XMM-Newton} data.
We believe that this temperature difference is caused by a combination
of instrumental effects including: the low sensitivity
of the ACIS-I instrument below 0.6 keV relative to the {\em XMM-Newton}
cameras, the complex temperature structure in the
ISM, systematic uncertainties in the low-energy quantum
efficiency due to the build up of contamination (A. Vikhlinin, private communication)
\footnote{For further details, see the website $http://asc.harvard.edu/cal/Acis/Cal_prods/qeDeg/index.html$
for detailed discussion of the ACIS contamination}, and to
the complexities in the spatial non-uniformity of the gain of the
ACIS-I instrument at low energies, and systematic uncertainties in the
spectrum and normalization of the background..
We have adjusted the ARFs of the {\em Chandra} spectra of the ISM using the
{\em apply\_acisbs} script provided by the CXC and we still find temperatures
which are systematically higher than {\em XMM-Newton}.
We conclude that the ACIS-I instrument is not particularly
sensitive to temperature determination for plasmas with temperature below
about 0.5 keV.  We rely on the results of the analysis of the Cen A ISM
from the {\em XMM-Newton} data exclusively in the remainder of
this paper.  A 50\% higher temperature for the Cen A ISM would
change some of our final estimates of pressures, Mach numbers,
etc., for the dynamics of the radio lobe, but none of our basic conclusions.

\begin{figure}
\plotone{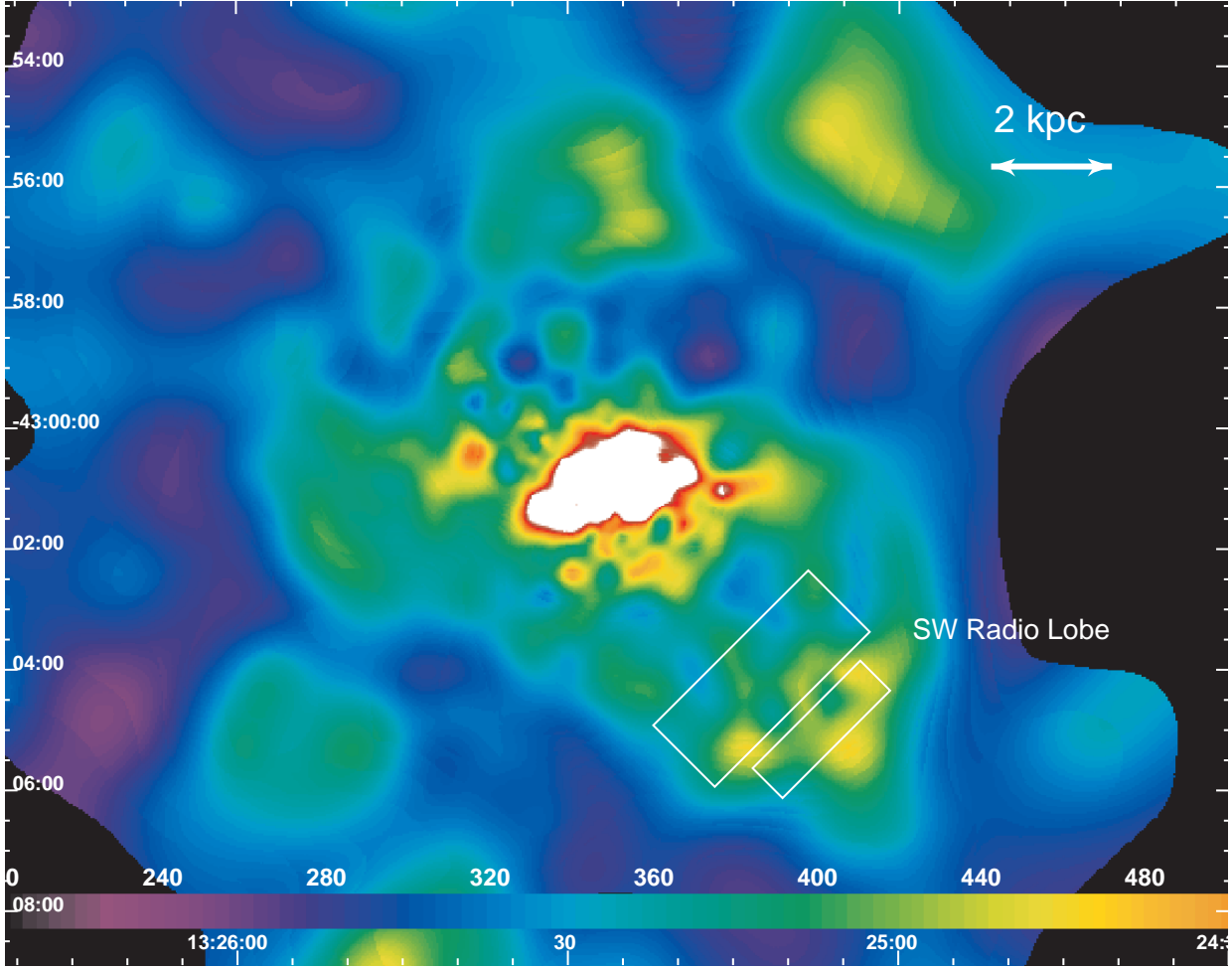}
\caption{Temperature map of the ISM from the {\em XMM-Newton} MOS data of Centaurus A.
A detailed explanation of the technique used to create this
map is contained in the text.  Note the
azimuthal temperature structure in the ISM. This is probably a result of
the merger, the interaction with the radio components, and the complex,
variable absorption.  The boxes overlaying the SW radio lobe correspond
the regions used for spectral analysis shown in Table~\ref{regtab}.}\label{tmap}
\end{figure}

Motivated partly by the complex X-ray morphology apparent in Figure~\ref{xmmrad}
and partly by the desire to confirm independently the
results of the spectral fitting of the {\em XMM-Newton} data,
we have created a temperature
map of the MOS data using the technique described in \citet{chu96}.
In this technique, images in five energy bands are created, and the
bin size of each cell is adaptively chosen so that the signal to noise ratio
is constant.  Simulated spectra are created using XSPEC, and the
temperature of each cell is determined via least squares fitting of the
data to these spectra.  The temperature map is shown in Figure~\ref{tmap}.
As can be seen from the figure, there is clearly azimuthal temperature structure in the ISM,
but the average temperature of the ISM is about 0.3 keV.  This independent
analysis is consistent with the temperature derived on the basis of
spectral fitting of the {\em XMM-Newton} data.
Much of the complex structure of the ISM, particularly in the central $2'$ of
the galaxy and to the NW and SE of the nucleus (i.e. perpendicular to the axis of the
radio components), is probably related to a recent galaxy merger and not due to interaction with the
relativistic plasma of the jet and lobes.  
A preliminary discussion of this phenomenon has been presented elsewhere
\citep{kar02}.  For our analysis below, we will use
the azimuthally averaged radial surface-brightness and temperature profiles.

\subsection{Gravitating Mass Distribution}

The best-fit index to the surface brightness profile
(see Table~\ref{bmodtab}), $\beta$=0.40$\pm$0.04 (the average of
the MOS and PN values), is
rather flat, but not inconsistent with recent {\em Chandra} observations
of the hot ISM of NGC 4697 \citep{sar01}, an isolated elliptical about twice as massive
as Cen A, and other early-type galaxies observed with {\em Einstein} \citep{for85,tri86}.
There is clearly some azimuthal structure in the surface brightness
profile, particularly to the NW of the nucleus, which is most
likely related to a merger event with a small spiral galaxy \citep{kar02,sch94}.

\begin{figure}
\plotone{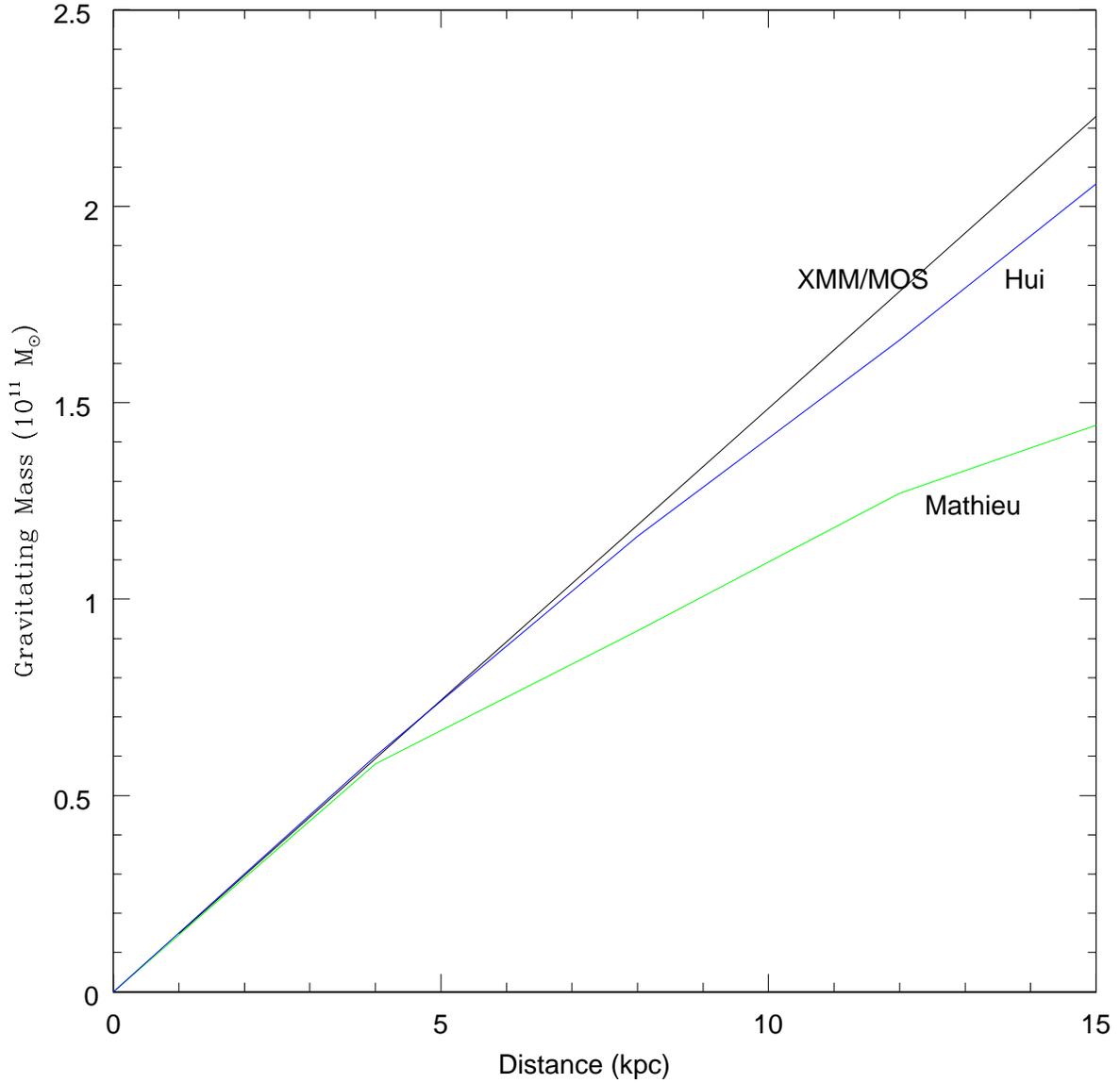}
\caption{Mass profile of Cen A derived from Equation~\ref{graveq} assuming
the gas is in hydrostatic equilibrium with the gravitating dark
matter potential (the black curve).  Two other estimates of the gravitating matter
based on analysis of the dynamics of planetary nebulae are
also shown \citep{hui95,mat96}.}\label{gravmass}
\end{figure}

Ignoring the surface brightness variations and assuming the gas is in hydrostatic equilibrium
with the dark matter potential of the galaxy, the total gravitating mass
within a radius $r$ of the nucleus is given by
\begin{equation}
M_G={\frac{-k_BT(r)r}{G\mu m_p}} \left(\left({\frac{d\ log\ n_H(r)}{d\ log\ r}}\right)+\left({\frac{d\ log\ T(r)}{d\ log\ (r)}}\right)\right), \label{graveq}
\end{equation}
where $n(r)$ and $T(r)$ are the radial distributions of the particle density and
temperature, respectively \citep{bah77,fab80}.  Assuming that $T(r)$=const=0.29 keV, and the particle
density is given by the $\beta$ model profile described in the previous section,
the total gravitating mass as a function of distance from the nucleus is shown in
Figure~\ref{gravmass}.  For comparison, two measurements of the gravitating matter
using planetary nebulae are shown as well \citep{hui95,mat96}.  These two different measurements
were made from the same data with different dynamical models.
We find that within 15 kpc of the
nucleus, the total mass of Cen A is $\sim$2$\times$10$^{11}$ M$_\odot$.

\subsection{Comparison of the X-ray Luminosities of the ISM and Point Sources}

The unabsorbed X-ray luminosity of the ISM within 10 kpc of the nucleus in Cen A 
is 7.71$\pm 0.54\times$10$^{39}$ ergs s$^{-1}$
(0.4-1.0 keV band).  This corresponds to an unabsorbed 
luminosity of 1.26$\pm 0.09\times$10$^{40}$ ergs s$^{-1}$
(0.1-10 keV band).  As with the $\beta$-model fit above, the uncertainty
in these numbers is dominated by the uncertainty in the background subtraction.
These numbers include an estimate of the effect of the
variable absorption of the dust lane, and the slightly higher temperature of the central $2'$.
The effect of the variable dust absorption was made by estimating an
average absorption of the dust lane using the extinction map
of \citet{sch96}.  This was converted to an equivalent column density using
the $N_H$/$A_V$ ratio determined from dust scattering haloes by ROSAT \citep{pre95}.
A scale factor was then computed using this value of $N_H$ in PIMMS, the
measured temperature of the central region, and the observed fraction of this inner
region that is obscured by the dust lane.
This X-ray luminosity is at the low end of the range of X-ray luminosities
for galaxies with the optical luminosity of Cen A ($M_B$=-20.4) (see Figure 2 of \citet{sul01}).
If the radial-surface brightness profile does not steepen for a large
distance beyond 10 kpc from the nucleus, this X-ray luminosity would only
represent a lower limit as the total X-ray flux would be dominated by the outer regions of the galaxy.

The integrated X-ray luminosity of all XRBs within $9'$ of the
nucleus of Cen A with $L_x >$10$^{37}$ ergs s$^{-1}$
(0.4 - 10.0 keV band) is 4.6$\times$10$^{39}$ ergs s$^{-1}$ \citep{kra01}.
Assuming a 5 keV thermal bremsstrahlung spectrum for the XRB population with galactic
absorption, this corresponds to an (absorbed) X-ray luminosity of 6.4$\times$10$^{38}$ ergs s$^{-1}$
in the 0.4 - 1.0 keV band.  The shape of the X-ray point-source
luminosity function (LF) below 10$^{37}$ ergs s$^{-1}$ is
somewhat uncertain because this is the approximate flux below which the observations
are complete and unbiased, but unless the LF
again steepens below 10$^{36}$ ergs s$^{-1}$, the integrated X-ray luminosity of sources
with $L_x <$10$^{37}$ ergs s$^{-1}$ contributes at most an additional few tens of percent to the
total X-ray luminosity of the point sources.
Below 1 keV, the XRBs contribute only $\sim$10\% of the integrated X-ray flux.
At higher energies, the XRBs contribute a significant fraction of the
total X-ray emission.

\section{The Radio Lobes}

\subsection{Imaging and Spectral Analysis}

\begin{figure}
\plotone{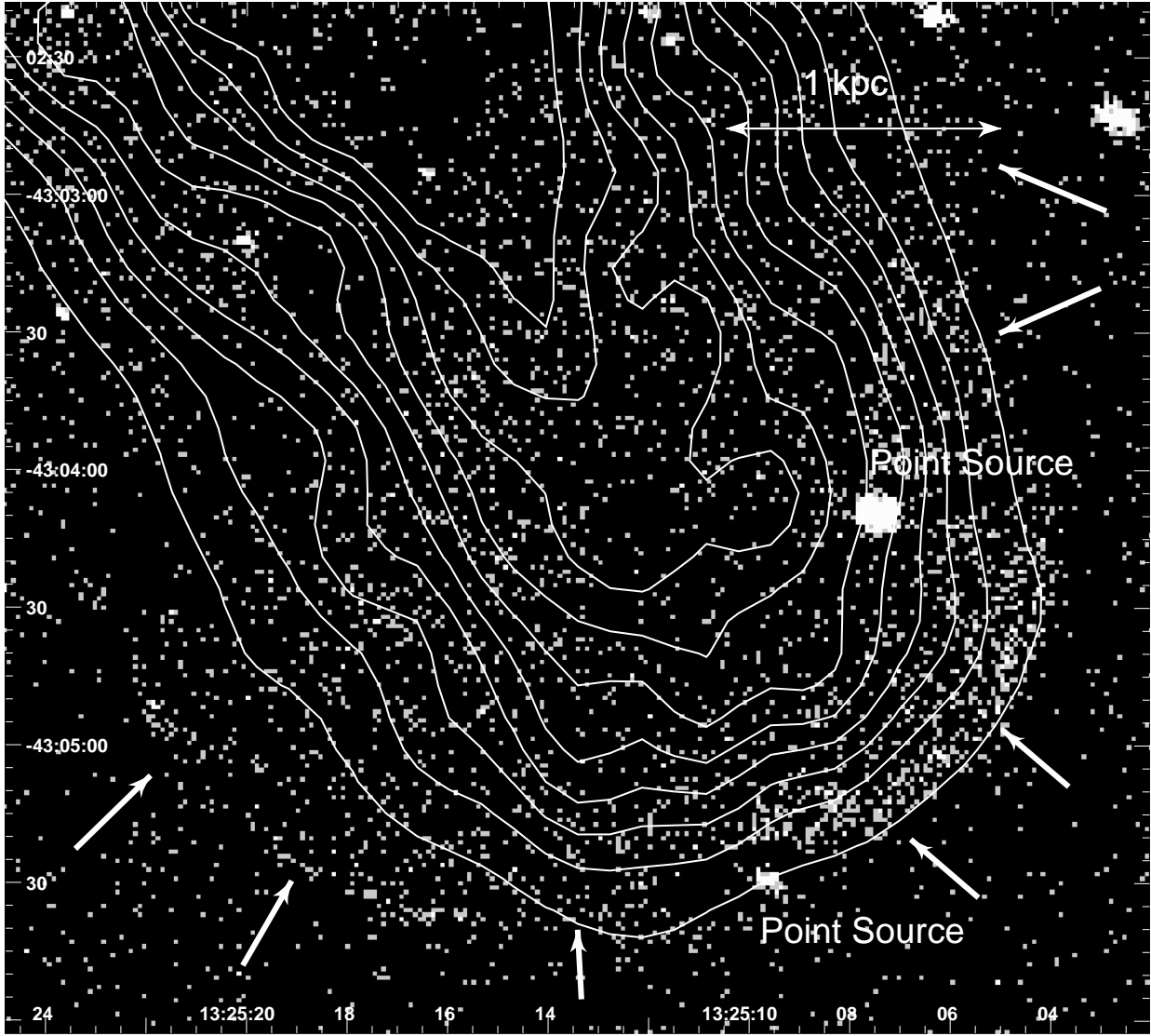}
\caption{Raw X-ray image (\chdr\ OBSID 00962 only) 
of the southwest radio lobe (0.5-2 keV band) with 13 cm contours overlaid.
Two point sources unrelated to the lobe are labeled, and several others are
visible in the NE portion of the image toward the nucleus.
The radio contours correspond to a flux density of 0.03 to 2 Jy/beam in ten logarithmic steps.
The radio beam is $30.45''$(RA) $\times$ $20.31''$(DEC). The arrows
indicate the approximate boundary of the X-ray shell.} \label{rawlobe}
\end{figure}

One of the remarkable features shown in Figures~\ref{chdimg} and~\ref{xmmrad}
is the X-ray enhancement coincident with the southwest radio lobe.
The relationship between the X-ray and radio emission of the SW lobe
is more clearly shown in Figure~\ref{rawlobe}, which
contains a raw {\em Chandra} image in the 0.5-2.0 keV
band with 13 cm contours overlaid.  The boundary of the X-ray
shell is highlighted by arrows.  
The bright X-ray enhancement is clearly visible along the edge 
of the radio lobe and appears to be contained within the lobe, although
the X-ray emission clearly lies beyond the NW and SE radio contours.
We argue that most or all of the X-ray enhancement actually
lies beyond the boundaries of the radio emission for two reasons.
First, the lobe is not in the plane of the sky (see below) and we
are seeing the features in projection.  Second, the radio map
has been considerably broadened by the $\sim 30''$ restoring beam, and
the bulk of the radio emission is therefore interior to the X-ray emission.

\begin{figure}
\plotone{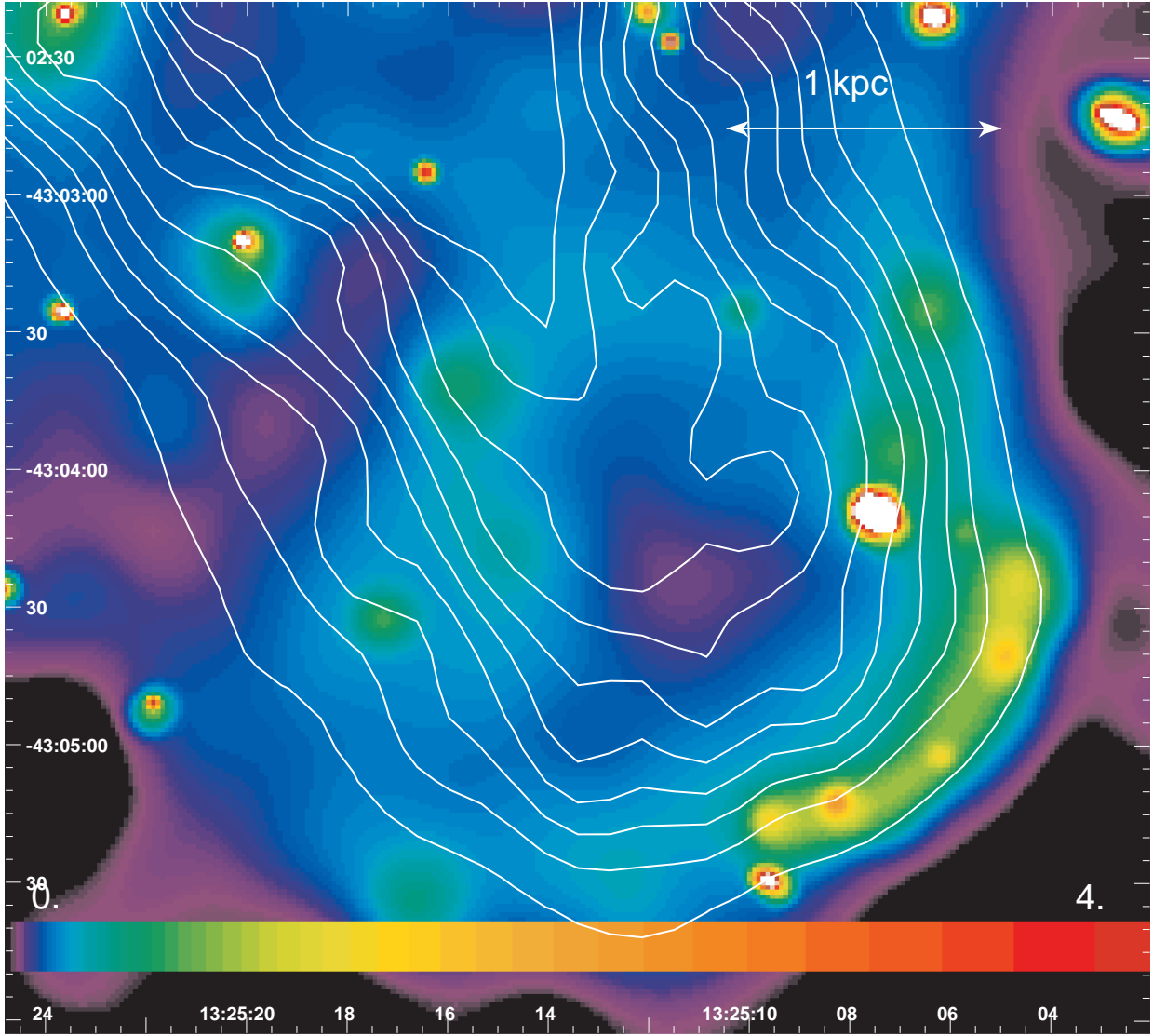}
\caption{Adaptively smoothed, exposure corrected, background subtracted X-ray image (\chdr\ OBSID 00962 only) 
of the southwest radio lobe (0.5-2 keV band) with 13 cm contours overlaid.
The X-ray surface brightness is indicated by the
color bar on the bottom of the figure in units of 10$^{-5}$ cts arcsec$^{-2}$ s$^{-1}$.
The largest smoothing scale of the X-ray data is $10''$.  
The radio contours correspond to a flux density of 0.03 to 2 Jy/beam in ten logarithmic steps.
The radio beam is $30.45''$(RA) $\times$ $20.31''$(DEC).  The knotty structures shown in
the enhancement are artifacts of the adaptive smoothing and
are not embedded point sources (see Figure~\ref{rawlobe}).} \label{swlobe}
\end{figure}

The diffuse structures of the X-ray shell are more clearly shown in Figure~\ref{swlobe},
which contains an adaptively smoothed, exposure corrected \chdr\
image of the southwest radio lobe in the 0.5-2 keV band with the 13 cm
radio contours overlaid.
There are significant differences between the X-ray and radio morphologies for the
emission that is clearly interior to the lobe (Figure~\ref{swlobe}).
The surface brightness of the X-ray emission in this region varies by a factor
of four.  The radio contours decrease monotonically from the center to the edge
of the lobe, whereas the X-ray emission shows several significant peaks and valleys not
seen in the radio.

We have divided the southwest lobe into two regions for spectral 
analysis to determine if there is any spectral difference between
the emission along the southwest edge of the radio lobe and the emission that
lies within the boundaries of the lobe.
The first region, which we refer to as the enhancement,
is a rectangular region along the southwest edge of the radio lobe.
The second region, which we refer to as the diffuse region, is
a rectangular region in the interior of the lobe.  The details
of these regions are summarized in Table~\ref{regtab}.
Region 1 is the
region along/beyond the southwest radio lobe and referred to as the enhancement
region in the text.  Region 2 is a region covering lobe emission but
excluding the bright foreground object, and referred to as the diffuse
region in the text.  The RA and DEC are the coordinates of the center of
the box, the Roll is the rotation angle anti-clockwise from north, and the dimensions of the box
are given by the width and length.  The regions are shown graphically on the
temperature map in Figure~\ref{tmap}.

Unfortunately, the {\em XMM-Newton} observations were aligned such that
the MOS chip gaps were placed directly along the enhancement, and
one of the PN chip gaps was placed along the western edge of the lobe.  
The diffuse region does not intersect any of the gaps and provides a consistency test
among all the data sets, which avoids uncertainties due to event
filtering of events near chip boundaries or in the computation
of the appropriate response matrices.
The shapes and positions of these regions were chosen to
avoid contamination from the bright source
CXOU J132507.5-430401, which is believed to be a foreground
star \citep{kra01}.
A second point source, CXOU J132509.6-430530, located within
the enhancement region, was also excluded.
Two background regions for each data region were chosen to the W and the NW
of the nucleus at similar distances from the core as the
source regions.  These background regions were selected
because they are devoid of point sources in
the {\em Chandra} images.  

For both regions in the lobe, all five data sets (two \chdr~and three \xmm) were fit with
an absorbed power-law, and one- and two-temperature MEKAL thermal plasma
models in the 0.3 to 5.0 keV band.  The lower limit
was fixed at 0.3 keV because below this energy the
\xmm~response is somewhat uncertain, and the ACIS-I instrument
has little response.  Above 5.0 keV, the source
flux falls, and the instrumental background
and contamination from the bright, heavily absorbed nucleus
become increasingly important.  We intially performed the fits
with the column density, $N_H$, and the elemental abundance (for thermal
models) as free parameters.
It was found that best-fit column density was always consistent with the
galactic value (7$\times$10$^{20}$ cm$^{-2}$ \citep{sta92}).
The elemental abundance was poorly constrained because of the relatively high
temperature.  We therefore decided to fix the column at the
galactic value and the elemental abundance at 0.4 times solar.  

The results of the spectral fits are summarized in Table~\ref{specfit}.
All fits were done independently, and all data were binned to
a minimum of 30 counts per bin.  All errors are 90\% confidence for
one free parameter.  The rate is the background subtracted count rate in
units of 10$^{-2}$ cts s$^{-1}$.
The indication $UC$ in the error column signifies that
that parameter is not meaningfully constrained.  The parameter $R_{thermal}$ in the
two temperature MEKAL model fits is the ratio of the best-fit emission
measures.  This parameter demonstrates that even in the cases where a second
temperature component improves the fit, the emission measure of the lower-temperature
component is small and that this second component
contributes negligibly to the pressure.
No parameters of the \chdr~spectral fits of region 2 are included as
the parameters are not constrained in a useful way due to the low surface
brightness.  The \xmm~spectra have several times
the number of counts that the \chdr~spectra have, and
the \xmm~fits are generally better (lower $\chi ^2_\nu$)
and have smaller error bars.  We therefore use the \xmm\ fits to determine the
spectral parameters (i.e. the temperature or power-law index), and use the \chdr~result
as an independent confirmation.  
As can be seen from Table~\ref{specfit}, the thermal and the power
law models both provide an adequate description of the data.  There is
no reason to favor one model over the other on the basis of spectral
fitting alone.  

For reasons we outline below, however,  
the thermal model is more physically plausible.
The addition of a second thermal component to the MEKAL fits generally improves
the quality.  We attribute this to a complex temperature
distribution within this feature.  It could indicate some spatial non-uniformities
in the distribution of the hot ISM around Cen A, but we consider this
less likely.  The temperature of this
second component is generally poorly constrained.
Although the addition of a second temperature component improves the fits,
this component is roughly an order of magnitude lower in temperature and a factor of
a few lower in emission measure than the hotter component.
This second component contributes little to the pressure of this feature.

Given the spectral similarity between the `enhancement' region and the
`diffuse' region, we will treat
the entire lobe as a single feature and make no distinction between the
emission interior to the lobe and the enhancement along the edge.  There is
some marginal evidence (see Table~\ref{specfit}) that the temperature
of the material that appears to be interior to the lobe is somewhat
cooler ($\sim$ 10\%) than that along the edge enhancement.  This is
what one would expect for the supersonic lobe expansion model, 
our preferred model, but for simplicity we will
ignore this small difference in the discussion below.  We will use the
results of the single-temperature fits of the enhancement region
for all estimates of density and pressure when discussing the thermal model.
Because of complications in the point-source removal and the
unfortunate alignment of the chip gaps in the {\em XMM-Newton} data described above,
the conversion of X-ray flux to particle density was determined by using
the spectral parameters of the {\em XMM-Newton} data, but the count rate
from the second (OBSID 00962) {\em Chandra}/ACIS-I observation for the entire lobe.  
We determined the count rate in the {\em Chandra} data set in a $1.84'$ radius circle
centered on the X-ray lobe in the 0.5 to 2.0 keV band with all point sources removed.
Background was estimated from an identical region approximately the same
distance from the nucleus with all point sources removed as well.
Assuming a temperature of 2.88 keV (the average of the eight single-temperature
MEKAL fits in Table~\ref{specfit}), the
observed {\em Chandra}/ACIS-I count rate (7.54$\times$10$^{-2}$ cts s$^{-1}$ in the 0.5 to 2.0 keV band)
corresponds to a flux of 9.6$\times$10$^{-13}$ ergs cm$^{-2}$ s$^{-1}$ (unabsorbed) in the
0.1 to 10.0 keV band, and an X-ray luminosity of 1.41$\times$10$^{39}$ ergs s$^{-1}$
at the distance of Cen A.

\subsection{Interpretation}

X-ray emission from the jets, hotspots, and radio lobes of radio
galaxies has been detected from several dozen sources with {\em Chandra}
\citep{wor01, har01, dhr00, bru02}.  In most of these cases, 
the X-ray emission has been attributed to non-thermal process (i.e. inverse-Compton
scattering of a variety of seed photons or synchrotron radiation) from a population of
relativistic electrons \citep{har02}.  
Although we cannot formally reject a non-thermal hypothesis for Cen A on the
basis of the spectral analysis alone, physical arguments
outlined below lead us to the conclusion that the X-ray emission from the southwest radio 
lobe of Cen A is most likely thermal in origin.
The interaction between the southwest radio lobe and the X-ray enhancement
of Cen A is most likely to be more closely analogous to the radio plasma/ICM interactions
recently observed by {\em Chandra} in early-type galaxies
\citep{fin01,jon02} and clusters of galaxies \citep{mcn00,dav01,fab00,bla01r} than
to non-thermal sources of X-ray emission.
Cavities and X-ray shells created by the inflation/expansion of `bubbles' of radio 
plasma in the ICM of galaxy clusters have been investigated
theoretically by several authors \citep{rey01,brg02m}.
These 'bubbles' are thought to be the backflow material from the
propagation of the powerful jets of radio galaxies through the ISM/ICM.  
The 'shells' of enhanced X-ray emission are due to the compression and
shock heating of the ambient ISM/ICM after it passes through the bow shock.
The Cen A X-ray enhancement around the southwest radio lobe appears to be the visible result of
a smaller-scale, lower-power (FR I jet and galactic ISM vs. FR II jet and ICM)
example of this radio plasma/thermal gas interaction.

We consider in detail three possible models for the origin of the X-ray
enhancement around the southwest radio lobe, two non-thermal and one thermal.  
First, we address the possibility that this emission is due to inverse-Compton
scattering from the relativistic electrons of
the radio lobe.  The most significant source of seed photons is from
the CMB, but unless the lobe is far from equipartition, it is
unlikely that a significant fraction of the X-ray emission
originates from this mechanism.  Second, we investigate the possibility that the
emission is synchrotron radiation from a population of ultra-relativistic
electrons in the radio lobe.  Because of the short lifetime of these
particles, we reject this hypothesis as well.
Finally, we explore the possibility that the emission originates from
a thermal plasma that surrounds the radio lobe.  In this model,
which we consider the most plausible,
the expansion of the lobe, energized by the counter jet, has compressed and shock-heated
the ambient hot ISM.  A partial shell or cap of plasma surrounding the lobe and rotated with respect to
our line of sight would naturally give the edge brightened appearance.

\subsubsection{IC scattering of CMB photons}

First, we consider the possibility that the X-ray emission is
due to inverse-Compton scattering of CMB photons off radio-synchrotron-emitting
relativistic electrons.  There are several significant problems with this model, given the observed
morphology and spectrum, that make it implausible.  First, the spatial morphology of the
X-ray emission argues strongly against this model.
The most prominent part of the detected X-ray emission, the enhancement ahead
of the lobe, has no detected radio counterpart down to at least
a frequency of 327 MHz \citep{sle83}.
The IC-scattering hypothesis
would require a large population of low-energy ($\gamma\sim$2000) electrons without
any extension to higher, observable energies.
The ratio of X-ray to radio flux in the interior of the lobe and along
the enhancement are significantly different, therefore implying
a significant difference in the non-equipartion conditions or energy spectral index of the relativistic
electrons in these regions.  The synchrotron frequency for the electrons responsible for the
IC scattering of the CMB photons into the X-ray band
is 80 MHz assuming the equipartition magnetic field of 13 $\mu G$ \citep{bur83}.
Emission at this frequency from the narrow shell would be
difficult to detect, but there is no plausible reason
to expect that there is a large population of $\gamma\sim$2000 electrons while
the presence of $\gamma\sim$10$^{4}$ electrons is ruled out by the lack of
detection of the shell by the VLA or the ATCA.

It is conceivable that a significant fraction of the diffuse emission in the interior of the lobe
is due to IC scattering of CMB photons, and that the enhancement along
the edge of the lobe originates in a different mechanism.  Such IC scattered
X-ray emission has been observed in the radio lobes of several powerful radio
galaxies \citep{fei95,tas98,hrd02b}.
To investigate the importance of this mechanism, we used
the code of \citet{har98} to determine
the magnetic field strength that would be necessary to produce the
observed X-ray flux density given the known radio source properties.
We modeled the electron spectrum in the lobe as a power-law of the
form $N(\gamma) {\rm d}\gamma = N_0 \gamma^{-p} {\rm d}\gamma$ with
$p=2$, $\gamma_{\rm min} = 100$ and a value of $\gamma_{\rm max}$
determined by the radio data at 2.3 and 8.4 GHz. We measured the
X-ray flux density (82 nJy at 1 keV) from 
a circular region $1.67'$ in radius covering most of the
outer or southern half of the radio lobe, but excluding the enhancement
along the edge of the lobe.  The X-ray background was estimated from a region of the
observation away from the lobe but at approximately the same distance from the nucleus.
We find that the magnetic field must be an order of magnitude below the minimum-energy value
(determined assuming a tangled field in the lobes) of 20 $\mu$G, if
the IC process is to produce all of the observed X-ray emission from
this region.  This is a much larger departure from equipartition than is
observed in other sources. If instead the magnetic field strength
has its equipartition value, the IC process can produce only $\sim
2$\% of the observed X-rays.

Other sources of seed photons are even more implausible than the CMB.
The synchrotron-self Compton (SSC) mechanism can be rejected immediately
as the density of radio photons is lower than that from the CMB.
Optical photons from either the nucleus or the stellar component
are two other possibilities.  Both are unlikely.
In the unified model of AGN, Cen A, as an FR I galaxy, is considered to
be a misdirected BL Lac \citep{urr95}.  In this case,
there may be a considerable flux of beamed
optical photons emitted from the nuclear region \citep{fos98}.  
To upscatter optical photons into the X-ray band,
the electrons must have $\Gamma = ({{\nu _X}/{\nu _{opt}}})^{1/2} \sim 100$.
The typical optical luminosity of beamed photons from an FR I radio
galaxy/BL Lac is
$\sim$10$^{42}$ ergs s$^{-1}$ \citep{pad91,chi01,har00}.  If all of these
photons are emitted from the nucleus into the solid angle
subtended by the southwest radio lobe, the energy density at the
edge of the lobe is $\sim$10$^{-13}$ ergs cm$^{-3}$.  It would
require approximately 3$\times$10$^{62}$ electrons with $\Gamma =$100
to produce the observed X-ray luminosity via IC scattering.
The total energy of these electrons, $\Gamma N_e m_e c^2$, is
$\sim$10$^{59}$ ergs, several orders of magnitude larger than
the equipartition energy of the lobe.  Assuming they are
distributed uniformly throughout the outer or southern half of the radio 
lobe, the pressure of these relativistic electrons would be
$\sim$3$\times$10$^{-8}$ dynes cm$^{-2}$, several orders of
magnitude larger than the equipartition pressure of the lobe.
The existence of such a large number of relativistic electrons
is improbable.  The energy density of optical starlight at the
edge of the lobe is of the same order of magnitude as that estimated for
the beamed nuclear source \citep{fei81}.  We therefore reject the
IC scattering hypothesis.

\subsubsection{Synchrotron Radiation}

X-ray synchrotron emission has been detected from a number of low-power jets by
\chdr, and may be a common feature of these jets \citep{wor01}.
Such a model was invoked to explain the X-ray
emission associated with the forward jet of Cen A \citep{fei81,kra02}.
The main difficulty with a synchrotron model in the context of the
Cen A radio lobe is, however, the short lifetime of the particles.
Defining the lifetime of the particles as the timescale for the
upper synchrotron cutoff frequency to become equal to the observing
frequency, the particles will lose their energy
on the order of tens of years in the equipartition magnetic
field \citep{bur83}. For comparison, the light travel time
across the lobe is approximately ten thousand years.
This would imply that the particles must be re-energized
hundreds or even thousands of times in the process of traveling
through the lobe.  

The X-ray and radio knots in jets are commonly
thought to be the sites of shocks where particle reacceleration
occurs.  No such knotty structure is seen in the
X-ray surface brightness of the radio lobe (Figure~\ref{chdimg}).
At the resolution of \chdr, the forward jet has a complex
morphology on scales from tens of parsecs to kiloparsecs \citep{kra02}.
No such structure is seen in the X-ray morphology of the southwest lobe. 
It is conceivable that there are thousands of small-scale knots which
are unresolvable with \chdr\ but give the appearance of more or
less uniform emission in the interior of the lobe.
We consider this unlikely in the absence of a distribution which includes
large knots.  The X-ray morphology strengthens this argument.
The projected length of the X-ray enhancement along the edge of the lobe
is quite small ($\sim$100 pc) compared with the
size of the lobe.  It is difficult to see how a synchrotron model with thousands of
unresolved knots would produce such a sharp, well defined feature
exactly along the edge of the radio lobe.
Finally, if the X-ray emission were due to synchrotron radiation,
the enhancement at the edge of the lobe would be marginally
detectable in the optical unless the spectrum flattens significantly
between the X-ray and optical.  Such a feature has not been seen.

\subsubsection{Thermal plasma}

As a third scenario,
we consider the possibility that the emission originates
in a partial shell or cap of thermal plasma that surrounds the radio lobe.
We suggest that little or none of the emission originates within the lobe, and that the
lobe is surrounded by a shell of plasma.  All of the X-ray
emission that appears to be within the lobe actually
originates in a sheath or cap of hot plasma that surrounds
the lobe.  That is, the emission that appears to be within the lobe
is actually in front of and behind the lobe along our line
of sight.  Emission from a thin cap of plasma rotated to
our line of sight would appear as an edge-brightened shell
and thus could take on the morphology of the detected emission.

It is reasonable to expect that the southwest radio lobe
and the surrounding plasma shell are rotated to our line of sight
at a complementary angle to that of the forward jet.
Various estimates have been made of the angle to the line of sight
of the forward jet, and all are highly uncertain.  
The most recent estimate of 50$^{\circ}$-80$^{\circ}$
is based on the VLBI brightness ratio of the
milliarcsecond forward jet and the counter jet \citep{tin98}.
Other estimates typically range from 55$^{\circ}$-70$^{\circ}$
\citep{gra79,duf79,ski94,jon96}.  In this work, we will assume
a value of $\theta _{jet}$=60$^{\circ}$ for the angle between the forward jet and the
line of sight, the angle between the counterjet/southwest radio lobe is
therefore $\theta _{cjet}$=120$^{\circ}$.  None of our results or conclusions is
sensitive to this choice.

We calculated the surface brightness from an optically thin plasma shell of 
uniform density rotated at an angle $\theta _{cjet}$=115$^{\circ}$ to our line of
sight with inner radius $r_{inner}=r_1$
and outer radius $r_{outer}=r_1+r_2$, where $r_1$ is the radius of the lobe,
and $r_2$ is the thickness of the X-ray emitting shell.
This calculation is important for estimating the volume,
and therefore the density, of the shell as described below.

The parameter $r_1$ can be approximately measured from the width
of the radio lobe in a direction perpendicular to the counter jet direction.
The parameter $r_2$ can be estimated by measuring the thickness of
the enhancement along the direction of the counter jet.
The {\em Chandra} OBSID 00962 data were used for this
measurement because the lobe is close to the best focus of the telescope so
that broadening of this feature by the telescope PSF is insignificant.
The image of the lobe in this observation is considerably sharper than in
the OBSID 00316 observation, where the lobe is positioned at the
very edge of the FOV.  We find the enhancement width to be $\sim 15''$ (FWHM) on the sky by fitting
a Gaussian plus second-order polynomial (to account for background and/or
more extended emission in the interior of the lobe) to the peak.  The effect of projection
on the sky was then removed by comparing the measured width
with the simulated width for a range of parameters $R$.
For our assumed counterjet angle ($\theta _{cjet}$=120$^{\circ}$),
we estimate a compression of approximately 8:1 (i.e. $R$=0.875) based on a comparison
of the measured width of the projected enhancement with that in our
calculated images.  The radius of the radio lobe is $\sim 1.84'$, so
the deprojected thickness of the shell is $\sim$16.6$''$ or $\sim$281 pc.
Note that this observed compression ratio is consistent with the expected distance between
the bow shock and the contact discontinuity if the lobe is expanding supersonically
\citep{ale02,nul02}.

We have based our estimate of the volume primarily
on the width of the enhancement along the edge of the lobe.
The sides of the shell in our preferred scenario will probably be thicker than at the
enhancement along the leading edge of the lobe.  
Systematic uncertainty in the thickness of the shell or oversimplifications of the
assumed geometry will not change any of our general conclusions about the
nature of the shell.
The particle density as a function of the compression parameter $R$ is
given by
\begin{equation}
n_H\sim V^{-0.5}\propto (1-R^3)^{-0.5}, \label{compreq}
\end{equation}
and is therefore not a strong function of the assumed thickness of the shell (unless
the shell is much thinner than we have assumed in which case the density would
be even higher).  If the material were uniformly distributed throughout the entire
southern half of the lobe, the shell density would only be a factor of two smaller.  
Given the uncertainty in $\theta_{jet}$, we estimate the uncertainty on
the volume to be on the order of 30\%.

In Table~\ref{presstab} We summarize the temperatures, the densities, and the pressures
of the emission in the shell, the ISM, and of the radio lobe (equipartition pressure from
the middle of the SW lobe taken from \citet{bur83}).
As can be seen from the table, the shell is
greatly overpressurized relative to both the ambient ISM and
the radio lobe, and the lobe is greatly overpressurized relative
to the ISM.  We infer that both the
lobe and the shell must be expanding supersonically
(relative to the ISM) and that the shell is
confined by the ram pressure of the expansion.
We hypothesize that the shell of hot plasma
around the lobe is the result of the advance of an unseen,
supersonic (and possibly now extinct) counter jet and the displacement and compression of the
ISM as the radio lobe inflates.  The current existence of a counter jet
is not necessary, only that it existed sometime in the past to initiate
the flow.  The overpressurization
and thin extent of the shell require the expansion to be supersonic
because otherwise the
shell would have dissipated on a time scale smaller than the expansion of the lobe.
In particular, the sharp boundary of the shell would
dissipate in a timescale much smaller than the expansion of the lobe
if the expansion were subsonic or transonic.
If this shell is indeed due to the supersonic expansion of the
radio lobe into the ISM, the simplest interpretation
of the thermodynamic properties of the gas
would be that we are directly observing the bow shock
across which the Rankine-Hugoniot (RH)
conditions are formally met \citep{lan90,spi98}.  In our case, however,
we cannot be directly observing the bow shock
because the density contrast between the shell and the
ISM is considerably larger than the factor of $\sim$4 (for $\gamma$=5/3) expected
for a strong shock.

Our scenario is a bit different than a shock wave propagating
into a uniform density medium however.
Both the gas density and pressure of the ISM are
falling rapidly ($\rho_{ISM}\sim r^{-1.2}$) as the lobe expands away from the nucleus,
and the lobe/shell system may be expanding self-similarly.
Such a scenario has
been investigated theoretically by several authors in the context of much more powerful
FRII jets propagating through the ICM of clusters of galaxies \citep{cla97,hei98,kai99,rey01,ale02}.
If the density gradient of the ICM is steep enough,
the density contrast between the shocked shell and the
ambient ISM/ICM will be considerably larger than 
that expected on the basis of the RH conditions \citep{hei98,kai99}.  It will appear as if
the gas was further compressed as it crossed the bow shock.  In fact, the
thermodynamic state of the gas in the shell depends on its past history
as it expanded through the denser regions of the ICM.
We suggest that such a model can explain the observed X-ray
emission around the southwest radio lobe of Cen A.  We note that
this problem has some similarities with that of a supernova explosion
expanding into the ISM \citep{sed59,che74,spi98}.

\begin{figure}
\plotone{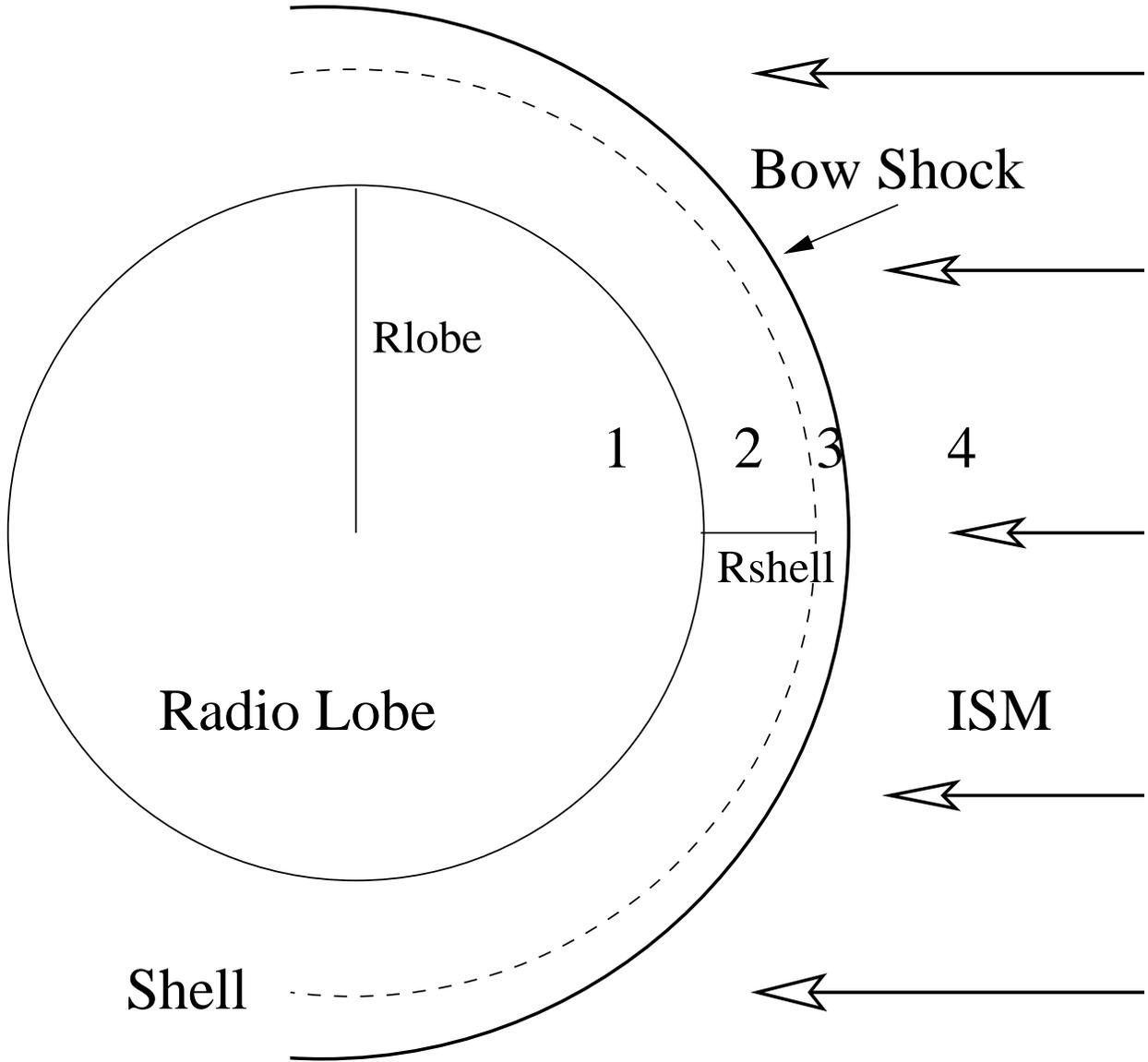}
\caption{Schematic diagram of the four region model (not to scale).  Region 1 is
the radio lobe, region 2 the observed X-ray enhancement region,
region 3 is a physically thin layer where the RH shock
conditions are met, and region 4 is the ambient ISM.}\label{shellmod}
\end{figure}

A detailed hydrodynamic simulation is required to fully interpret and
understand this phenomenon and will be the subject of a future paper, but we can
make some general quantitative statements about the evolution and
energetics of the lobe.  We have created a four-region model of
this system as shown in Figure~\ref{shellmod} in the rest frame of the
lobe.  Region 1 is the radio lobe,
region 2 is the X-ray shell, region 3 is a thin, hot boundary layer between the ISM and the X-ray shell
where the RH conditions are met,
and region 4 is the ISM.  The subscripts 1,2,3, and 4 will be used to
denote the thermodynamic variables and the thicknesses of each
region.  The lobe is expanding into the ISM with
velocity v$_{exp}$=v$_4$.  We assume that the lobe is expanding supersonically
into the ISM so that the RH shock conditions apply across the interface between regions
3 and 4 (i.e. v$_{exp}$=v$_4 >> c_4$ where $c_4$ is the sound speed in region 4).
We make the further assumption that region 3 is thin compared to the other regions (i.e. $r_3 << r_2$)
so that the observable emission from the shell is dominated by region 2.
Under these assumptions, we can determine the thermodynamic state of the gas in region 3
and estimate the expansion velocity of the lobe.

The temperatures, pressures, and densities of the gas in regions 2 and
4 are known on the basis of the X-ray spectral analysis presented above.
Since the RH conditions apply across the interface between regions 3 and
4, the density and velocity in region 3 are given by $n_{H3}=4\times n_{H4}=$6.8$\times$10$^{-3}$ cm$^{-3}$
and v$_{4}=4$v$_3$.
Balancing the thermal pressure of the gas in region 2 with the thermal
and ram pressure of the gas in region 3,
$T_3\sim$6.5 keV.  Region 3 therefore represents
a region of high temperature between the visible X-ray shell and the ISM.
Given the high temperature and low density (relative to region 2),
region 3 is in practice indistinguishable and/or unobservable 
in the {\em Chandra}/{\em XMM-Newton} band.  The ratio of the temperature
of the ISM to that of the boundary region is given by \citep{lan90}
\begin{equation}
{\frac{T_3}{T_4}}={\frac{2\gamma M^2 - \gamma +1}{\gamma +1}}{\frac{\rho _4}{\rho _3}},
\end{equation}
where $M$ is the Mach number of the lobe expansion into the ISM (i.e. v$_{exp}$/$c_4$).
We find an expansion velocity for the lobe of approximately Mach 8.5, or $\sim$2400 km/s.
This confirms our assumption in the previous paragraph that the lobe is
expanding supersonically into the ISM and the RH shock conditions are appropriate.
This analysis implicitly assumes that the lobe and the bow shock are
expanding at the same velocity.  If we add an additional constraint
and force the lobe/bow shock expansion to be self-similar, the derived
shock temperature would be somewhat higher ($\sim$8 keV).

The pressure of the plasma in the shell (region 2) is 
an order of magnitude larger than the equipartition pressure of the radio
lobe (region 1).  It is reasonable to Assume that these two components are
in approximate pressure equilibrium.  This implies that
there must be an additional component providing pressure support in the
radio lobe, perhaps protons or lower energy electrons, for the shell not
to destroy the lobe on the sound crossing time.
Such large deviations from the equipartion conditions have
been inferred for other radio sources \citep{har98b},
but not on such small scales as we see in Cen A.

The temperature between the shock temperature (6.5 keV) and the
measured temperature of the gas (2.9 keV) in the shell indicates that there
must be significant cooling in the shell as the material flows away from
the bow shock.  Neither radiation nor thermal
conduction are likely to be important.  The radiative timescale for
the material in the shell is on the order of 10$^{8}$ yrs, whereas the dynamical
timescale of the lobe is on the order of 10$^{6}$ yrs.  Therefore radiative
cooling cannot be significant.  Likewise, the timescale for thermal conduction to the ISM
must be longer than the dynamical timescale because the boundaries of the
shell are sharp.  This, and the inferred adiabatic evolution of the
shell described below, imply that the thermal conductivity in the shell
and between shell and the ISM is considerably lower than the canonical
value of \citet{cow77}.  A similar suppression of transport
processes has been observed in the ICM of galaxy clusters \citep{vik01}.
Thermal conduction between the shell and the relativistic plasma in the 
radio lobe would heat the shell, not cool it.
The only other possibility is adiabatic expansion.  Assuming that the lobe and shell
expand self-similarly, the volume of the shell increases as the lobe inflates so
that the temperature of the gas must decrease.

The adiabatic evolution of a shell of shocked gas created by the supersonic
inflation of a radio lobe in a $\beta$-model atmosphere has been
studied by \citet{ale02}; see also \citet{dys80}.  As the
lobe and shell expand self-similarly, new material is constantly
added to the shell, but the material currently in the shell cools adiabatically
as the volume of the shell increases.
\citet{ale02} finds that if the density of the
atmosphere into which the lobe/shell is expanding falls
faster than $r^{-1}$,
a temperature and density gradient between the bow shock and
the contact discontinuity will be formed with the coolest material
lying just above the contact discontinuity.  
This material will have the largest
X-ray emissivity, so that the spectral parameters we derive are representative
of this region of the shell.  This scenario is therefore
qualitatively consistent with our inferred temperature gradient
between the shock temperature ($T_3$) and the temperature of the
shell ($T_2$).  Using the formalism of \citet{ale02} and assuming
$\rho_{ISM} (r) \sim r^{-1.2}$ (see Section 3.1), their model predicts a factor of $\sim$1.4 temperature
difference between the material along the contact discontinuity and that
just behind the bow shock.
A larger temperature gradient would be created if the value of
$\beta$ were larger, but the lobe must have progressed through an
atmosphere with a steeper density
gradient ($\rho_{ISM} (r) \sim r^{-\alpha}$ with $\alpha\sim$1.7-1.8)
to account for the observed temperature gradient.
Given the formal uncertainties in the
thermodynamic parameters and assumptions (i.e. uniform density and
temperature of the shell, using the average $\beta$-model to describe
the ISM when there are large azimuthal asymmetries in the surface
brightness, the unknown temporal history of the jet powering
the lobe, the complex environment within 2 kpc of the nucleus and
possible complications due to cold or warm gas remaining from the
merger, etc.) we do not consider the discrepancy between the
predicted and measured density gradient to be significant.

If the material in the shell is to behave adiabatically, the thermal
conduction must be effectively suppressed.  The heating timescale, $\tau _h$,
of the material in the shell is given by \citep{cow77,ale02}
\begin{equation}
\tau _h \sim 0.03 \tau_s({\Delta r}/\lambda _e)/\eta,
\end{equation}
where $\tau_s$ is the sound crossing time of the shell, $\Delta r$ is the thickness
of the shell, $\lambda _e$ is the electron mean free path, and $\eta$ is
the suppression factor.  For the shell around the SW radio lobe, the
canonical heating timescale ($\eta$=1) is $\sim$2$\times$10$^{5}$ yrs, approximately
an order of magnitude less than the dynamical timescale of the lobe.
Therefore, the thermal conductivity must be suppressed by a factor
of 100 or more for the shell to remain adiabatic and support the
inferred temperature gradient.  The swept up magnetic field, while
not dynamically important in the shock for reasonable assumptions
of the ambient field strength, could effectively suppress thermal
conduction in the shell if shear flows are present in the shell
(e.g. if the shell is advancing more rapidly radially than laterally).

We have made a consistency check on this analysis by comparing the mass
of material in the shell with the mass of the ISM in the region
swept out and currently occupied by the lobe.
The total mass of material in the shell
is $M_{SH}$=$\rho _{SH}\times V_{SH}\sim$3.2$\times$10$^{6}$ $M_{\odot}$.
We estimate the mass of the ISM displaced by the expansion of the
lobe by integrating the $\beta$-model in a piecewise fashion (i.e. $\rho (r)$= const
for $r\le a$ and $\rho (r)\propto r^{-1.2}$ for $r > a$) over the conical
region currently occupied by the lobe and find $M_{ISM}\sim$2$\times$10$^{6}$ $M_{\odot}$.
To order of magnitude, the mass in the hot shell is consistent with
material swept up from the ISM.

One parameter commonly used to distinguish shock-heated, supersonically compressed gas
from adiabatically, subsonically compressed gas is the specific entropy,
$S\propto P/\rho^{5/3}$.  For shock-heated gas, the specific entropy will increase
whereas there will be no change for adiabatically compressed gas \citep{lan90}.
Using the values from Table~\ref{presstab}, the ratio of specific entropies
of the shell and the ISM at the edge of the shell is approximately unity.
As described above, however, the current thermodynamic state of the shell depends on the 
past history in a complex way.  It is probably more relevant to
compare the current state of the shell with the density (and temperature)
of the ISM at a smaller distance from the nucleus.
This would imply that the specific entropy of the shell is indeed
larger than that of the gas and provides additional support for shock
heating of the shell.

As an alternative to the shock-heating of the hot phase
of the ISM, we consider the possibility that much cooler gas has been
shock-heated by the supersonic expansion of the lobe.
Considerable amounts of cold gas (i.e. neutral or molecular, $T<10^4$K) are present in
Cen A, likely because of the merger with the spiral galaxy, although
this gas is distributed unevenly.  The HI observations of \citet{gor90}
and \citet{sch94} detected $\sim$8$\times$10$^{8}$ $M_{\odot}$ of
neutral hydrogen aligned along the dust lane (i.e. perpendicular to
the jet).  This mass estimate is actually a lower limit because
the mass in the central 2.5 kpc is uncertain due to absorption.
There is enough cold gas present in Cen A to account for
the mass in the shell if only a small fraction ($\sim$1\%) of
it was swept up and heated by the lobe.

The total mass of warmer (10$^{4-5}$K) gas in Cen A is more uncertain.  UV observations
of more massive early-type galaxies have placed a lower limit
on the mass of ionized gas of $\sim$10$^{6}$ $M_{\odot}$ \citep{tri91b,tri97}.
This estimate relies on a highly uncertain filling fraction and is
only an order of magnitude estimate at best.
Unless this is a considerable underestimate, however,
this mass of ionized material is probably not sufficient to
account for the material in the shell in Cen A.
The spiral galaxy may have had a considerable warm ISM component if
it were similar to the Milky Way, so that the amount of warm
gas in Cen A may be much larger than is typical for massive,
early-type galaxies.  In either case, 
the final temperature of the gas in the shell depends only on
the expansion velocity of the lobe, so that whether the shocked
material originates in the coronal gas or in cooler gas, the implied
expansion velocity of the lobe remains unchanged.

\subsection{Discussion}

\subsubsection{Dynamics of Jets and Radio Lobes}

The conventional paradigm suggests that the jets of FR I galaxies
like Cen A are expanding subsonically into the surrounding medium, whereas
the lobes of the more powerful FR II galaxies are expanding
supersonically \citep{ick92}.  The existence of the shock-heated shell
around the southwest radio lobe implies supersonic expansion which runs
counter to this prevailing view.
This suggests either that Cen A is different than most of the other
well studied FR I galaxies, or that its proximity allows us to observe details that
are not readily observable in more distant objects and the standard paradigm
is therefore incorrect.
The radio power of Cen A ($P_{1.4\ GHz}$=1.85$\times$10$^{24}$ W Hz$^{-1}$ \citep{coo65})
is comparable to those of well-studied FR I radio galaxies from the 3C sample.
Cen A is in some ways, however, not a typical FR I object
even though it is often referred to as the prototypical object of the class.
The jet/lobe morphology of Cen A is clearly
different than the `tailed twin jet' morphology \citep{lea93}
commonly associated with objects like 3C 31 \citep{hrd02,lai02} and 3C 449 \citep{fer99}.
The multiscale structure of Cen A also makes it distinct from the more common
bridged twin jet FR I sources such as 3C 296 \citep{lea91}.
One the other hand, if Cen A were at a distance comparable to these other FR I galaxies
($\sim$100 Mpc), the X-ray cap around the SW radio lobe would be
virtually undetectable.
The larger scale radio components of Cen A (e.g. the Northern Middle Lobe)
are likely to be evolving subsonically \citep{sax01}.

It is interesting that the X-ray morphology of the NE radio lobe is so different
from the SW lobe.  Based on a preliminary examination of our AO-3 ACIS-S observation of Cen A,
there is some evidence for a partial shell around the NE lobe, but with an
X-ray luminosity more than two orders of magnitude less than the shell around the
SW lobe.  These X-ray morphological differences
aren't too surprising given the very different radio morphologies of the NE
(edge brightened on one side) and southwest (center filled with filamentary structure) lobes.
This argues that the nature of the flows on kpc scales to the NE and southwest are fundamentally different,
and in particular that the environment must play a key role in the appearance
of the jets and lobes and in the overall dynamics of the flow.
There are larger scale
asymmetries in the radio emission from Cen A as well \citep{mor99}, so that the
past history of the radio source may play a key role in its current
appearance.

Prior to the launch of {\em Chandra}, X-ray shells associated with
the lobes of radio galaxies had been detected in only two objects, Perseus A \citep{boh93}
and Cygnus A \citep{cla97}.  Since the launch of {\em Chandra}, many
additional examples of the interactions between the ICM and the
jets of radio galaxies have been detected \citep{chu01,fin01,mcn00,dav01,fab00,bla01r}.
In all of these cases the shells are cooler than the surrounding ISM, in contrast
to what we have observed with Cen A.  In these other objects, radiative cooling
may be important, but it is more likely that the cool shells are due
to entrainment of lower temperature gas from the central regions of the
galaxy or cluster by the inflation of lobe.  It is clear that the
nuclear activity can have an important effect on the galactic or
cluster environment, and it is quite likely that the reverse is true
as well.

\subsection{Relationship between Nuclear Activity and Cooling Flows}

This observation of the southwest radio lobe of
Cen A provides the first clear demonstration
of the complex relationship between the hot ISM of early
galaxies and nuclear outflows from supermassive black holes at their centers.
Other recent {\em Chandra} observations have hinted at such a relationship \citep{jon02},
but we believe that this is the first definitive example where the nuclear
outflow is providing sufficient energy to heat the ISM.
The total thermal energy of the gas within
15 kpc of the nucleus is $\sim$1.8$\times$10$^{56}$ ergs if one assumes
the gas is isothermal with a temperature of 0.3 keV.  This is actually an underestimate
because the temperature rises somewhat in the central
2 kpc, but is accurate enough for our order of magnitude comparison.
The total thermal energy of the gas in the shell around the
southwest radio lobe is $\sim$4.2$\times$10$^{55}$ ergs,
a significant fraction of the energy in the ISM.  In its current
configuration, the energy of the shell is a few tens of percent of
the total energy in the hot ISM, and additional energy
will be deposited in the ISM as the lobe continues expanding.

The ultimate fate of the gas (and energy) in the shell is not clear.
It is possible that plasma in the shell will eventually come back into
thermal equilibrium with the ISM and settle back into hydrostatic equilibrium
with the gravitating dark-matter potential, effectively heating
the ISM.  On the other hand, the highly supersonic expansion of the lobe could drive the
heated plasma completely out of the galaxy into the Cen A group, or
even out of the group into the IGM.  The temperature of the gas
in the shell is also much too large to be gravitationally bound to
a galaxy as small as Cen A ($M_B$=-20.4 \citep{isr98}, $M$=2.2$\times$10$^{11}\ M_\odot$ from
Figure~\ref{gravmass}).
The ratio of the kinetic energy of the shell to the thermal energy
in the shell is $\sim$6.5.  Currently the kinetic energy of
expansion is a factor of a few larger than the thermal energy
of the shell, but it is likely that the shell is
decelerating.  It is clear that the expansion of the southwest radio lobe {\em can}
provide enough energy to reheat the hot ISM, although it is unclear
if it actually does.

The cooling time for the hot ISM in the central regions of elliptical
galaxies is $\sim$10$^{8-9}$ yrs, much less than a Hubble time.  Prior to the {\em Chandra}/{\em XMM-Newton}
era, it was therefore expected
that significant amounts of cool gas would be found in their central
regions.  Recent {\em XMM-Newton} RGS observations of Abell 1835 \citep{pet01},
M87 \citep{boh01,boh02,buo01} and NGC 4636 \citep{xu01} have not detected the spectral
signatures of this cool gas
and have cast serious doubts on the existence of
large amounts of cool gas expected on the basis of
observations with an earlier generation of X-ray observatories (see \citet{bri02} for a detailed
discussion).  A similar situation exists for cooling flows
in clusters of galaxies in that the large amount of cool gas ($T<0.2$keV) inferred from
ROSAT, {\em Einstein}, and ASCA observations is not being found
with {\em Chandra} or {\em XMM-Newton} \citep{mol01} grating observations.
In the absence of this cool gas, the ISM must be occasionally reheated because of the relatively
short radiative lifetime.

It has been suggested that there is a cyclical relationship
between the hot ISM of elliptical galaxies (and clusters of galaxies)
and nuclear activity \citep{cio97,cio01,chu01,bri02,jon02}.  
In these models, supermassive black holes (SMBH) at the center of
elliptical galaxies undergo intermittent outbursts 
as the hot ISM radiatively cools and accretes.
During the outburst, energy is transferred from the active
nucleus to the hot ISM via outflows that are observed as
jets and radio lobes in galaxies.  After the epoch of nuclear
activity ends, the host galaxies undergo a long period
of quiescence where the ISM in Cen A slowly cools.  Once a significant amount
of energy is lost from the inner regions of the galaxy, material
flows onto the SMBH and initiates another epoch of nuclear activity.
The implication is that all cooling-flow elliptical galaxies are radio galaxies,
or that they have been radio galaxies in their past.
This is still an open question.  It has been shown that
the host galaxies of FRI and FRII radio galaxies are drawn
from a random population of otherwise normal elliptical galaxies
\citep{sca01}.  The probability that an elliptical galaxy
is also a radio galaxy is a steep function of the host's
luminosity \citep{led96}.  The more luminous, and therefore more massive,
galaxies also tend to have the shortest cooling timescales
for the ISM at their centers, typically 10$^{8}$ yrs \citep{for85,can87,fab89r}.

\section{Summary and Conclusions}

We have presented the results from one {\em XMM-Newton} and two {\em Chandra} observations
of the hot ISM and radio lobes of the nearby radio galaxy Centaurus A.  We find that:
\begin{enumerate}

\item  The temperature of the ISM beyond 2 kpc from the nucleus is approximately
0.29 keV with a small
decrease in temperature as a function of distance from the nucleus.  The average
radial surface brightness profile is well described by a $\beta$-model with an
index of 0.39$\pm$0.04.  There is, however, some azimuthal structure in both the
temperature and surface brightness profile, most likely related to a recent
merger with a small spiral galaxy.

\item X-ray emission coincident with the southwest radio lobe is detected.
A sharp X-ray enhancement along the edge of the lobe is also observed.  
Based on arguments about the energetics, the spectrum, the electron
energy distributions, and the observed morphology, we reject a non-thermal
(i.e. synchrotron or inverse-Compton scattering) origin for the emission.
We model this emission as a shell or cap of hot plasma that surrounds the radio lobe.
The gas parameters of this shell were estimated using a simple
model for the observed width of the enhancement
along the edge of the lobe and an estimate of the angle made by the jet/counterjet
with respect to the line of sight.  

\item  Based on spectral analysis, the temperature and density of the shell
are much larger (factors of 10 and 11.8, respectively)
than the ambient medium.  The shell is enormously overpressurized and this
requires that the lobe and the shell are expanding supersonically into the
ambient ISM.  This conclusion is supported by the small linear extent of the
X-ray enhancement along the edge of the lobe.
We estimate a Mach number of about 8.5, or a velocity of 2400 km/s.

\item The density ratio between the material in the shell and that
of the ambient ISM is too large to be explained in terms of the canonical Rankine-Hugoniot
shock conditions.  We suggest that the appearance of the
additional compression is in fact due to the supersonic expansion of the lobe into
a medium with a steep pressure and density gradient.
Recent hydrodynamical simulations of the expansion of the lobes
of FR II galaxies into the ICM support this conclusion.
Hydrodynamic simulation is required to quantitatively understand this
phenomenon and will be the subject of a future paper.

\item The X-ray shell is also enormously overpressurized relative
to the equipartition pressure of the radio lobe.  An additional
component, perhaps protons or lower energy relativistic electrons,
must be providing pressure support in the lobe.

\item The amount of energy transferred to the ISM by expansion/inflation of the
radio lobe is a significant fraction of its total thermal energy,
demonstrating the complex, and perhaps cyclical, link between the
ISM on the one hand and nuclear activity and outflows on the other.

\end{enumerate}

This last point could provide a partial answer to one of the long-standing puzzles
in X-ray astronomy, namely why is there such a large variance in
the X-ray luminosity of early-type galaxies of a given optical
luminosity \citep{for85,esk95}.  The environment and depth of the
dark matter potential play a key role in this, but
the cyclical interaction between the ISM and nuclear
activity can contribute to this variance as well.
That is, the X-ray luminosity of an early galaxy
will depend where in this cooling/reheating cycle we happen
to be observing it. {\em Chandra} and {\em XMM-Newton} have observed
(and will continue to observe) a large number of early-type galaxies with
a wide range of nuclear activities and radio powers.  Perhaps once
a large enough sample has been observed and analyzed, a
trend can be developed to quantify this relationship.
Cen A is, in fact, considerably underluminous for its optical
luminosity which would support the idea that it is at the
end of its cooling cycle.  We are just catching Cen A as it
starts to reheat its ISM.

\acknowledgements

We would like to thank Mark Birkinshaw, Torsten En{\ss}lin, Sebastian
Heinz, and Francesco Minitti for
many stimulating and helpful discussions.
We would also like to thank Dan Harris and the anonymous referee for their
detailed comments about this paper.
This work was supported by NASA contracts NAS8-38248, NAS8-39073, the
Chandra X-ray Center, and the Smithsonian Institution.

\clearpage

\clearpage

\clearpage

\clearpage

\clearpage

\clearpage

\clearpage

\clearpage

\clearpage

\clearpage

\begin{table}
\begin{center}
\caption{Observation Log.}\label{obslog}
\begin{tabular}{|l|c|c|c|c|c|c|c|}\tableline
Instrument       & OBSID &  Date   & Exp Time (ks) & RA & DEC & Y Offset & Z Offset \\ \tableline
ACIS-I           & 00316 &  5DEC99 &  35.9 & 13:25:27.61 & -43:01:08.9 & $3'$ & $3'$ \\ \tableline
ACIS-I           & 00962 & 17MAY00 &  36.5 & 13:25:27.61 & -43:01:08.9 & $-3'$ & $0'$ \\ \tableline
{\em XMM-Newton} & 93650201 & 2FEB01 & 23.1 & 13:25:26.3 & -43:01:06 &   &  \\ \tableline
\end{tabular}
\end{center}
\end{table}

\clearpage

\begin{table}
\begin{center}
\caption{Summary of best-fit $\beta$ model parameters.}\label{bmodtab}
\begin{tabular}{|l|c|c|c|c|c|c|c|}\tableline
$\beta$ & kT (keV)     &  $n_0$(cm$^{-3}$) & $r_0$ (kpc) & Central Cooling time (yrs) \\ \tableline
0.40$\pm$0.04 & 0.29$\pm$0.3 & 4.0$\times$10$^{-2}$ & 0.5 & $\sim$10$^8$ \\ \tableline
\end{tabular}
\end{center}
\end{table}

\clearpage

\begin{table}
\begin{center}
\caption{Summary of regions used for spectral analysis.}\label{regtab}
\begin{tabular}{|l|c|c|c|c|c|}\tableline
Region & RA & DEC & Roll & Width & Length \\ \tableline
1 - Diffuse & 13:25:14 & -43:04:01 & 45$^\circ$ & $1.44'$ & $3.63'$ \\ \tableline
2 - Enhancement & 13:25:07 & -43:05:00 & 45$^\circ$ & $0.70'$ & $2.52'$ \\ \tableline
\end{tabular}
\end{center}
\end{table}

\clearpage

\begin{table}
\begin{center}
\caption{Spectral Parameters of Shell.}\label{specfit}
\begin{tabular}{|l|c|c|c|c|c|}\tableline
      & XMM-MOS1 & XMM-MOS2 & XMM-PN & Chandra - 00316 & Chandra - 00962 \\ \tableline
\multicolumn{6}{|c|}{Region 1 - Enhancement along/beyond the edge of the southwest lobe} \\ \tableline
Rate (10$^{-2}$ cts s$^{-1}$) & 2.30$\pm$0.12 & 1.63$\pm$0.11 & 7.75$\pm$0.26 & 1.67$\pm$0.08 & 1.62$\pm$0.14 \\ \tableline 
\multicolumn{6}{|c|}{Power Law Model} \\ \tableline
Index      & 2.04$^{+0.13}_{-0.14}$ & 1.80$^{+0.20}_{-0.20}$ & 1.81$^{+0.10}_{-0.09}$ & 1.85$^{+0.13}_{-0.12}$ & 1.83$^{+0.13}_{-0.14}$ \\ \tableline
$\chi^2_v$ & 1.73 & 1.11 & 1.02 & 1.50 & 0.97 \\ \tableline
\multicolumn{6}{|c|}{One Temperature MEKAL Model} \\ \tableline
$T_1$      & 2.83$^{+0.80}_{-0.50}$ & 2.62$^{+1.50}_{-0.70}$ & 3.70$^{+0.83}_{-0.70}$ & 3.4$^{+1.3}_{-0.7}$ & 3.4$^{+0.7}_{-0.7}$ \\ \tableline
$\chi^2_v$ & 2.92 & 0.49 & 1.24 & 2.31 & 1.32 \\ \tableline
\multicolumn{6}{|c|}{Two Temperature MEKAL Model} \\ \tableline
$T_1$      & 4.19$^{+2.9}_{-1.3}$ & 2.90$^{+2.10}_{-0.90}$ & 4.5$^{+1.7}_{-1.0}$ & 5.1$^{+4.0}_{-1.7}$ & 4.2$^{+2.8}_{-1.3}$ \\ \tableline
$T_2$      & 0.33$^{+0.10}_{-0.08}$ & 0.47$^{+UC}_{-UC}$ & 0.21$^{+0.07}_{-0.07}$ & 0.76$^{+0.20}_{-0.45}$ & 0.71$^{+UC}_{-UC}$ \\ \tableline
$R_{thermal}$  & 0.44 & 0.06 & 0.20 & 0.18 & 0.07 \\ \tableline   
$\chi^2_v$ & 1.12 & 0.50 & 1.03 & 1.46 & 1.30 \\ \tableline
\multicolumn{6}{|c|}{Region 2 - Diffuse emission coincident with/interior to the lobe} \\ \tableline
Rate (10$^{-2}$ cts s$^{-1}$) & 3.60$\pm$0.20 & 3.26$\pm$0.20 & 8.74$\pm$0.39 & & \\ \tableline
\multicolumn{6}{|c|}{Power Law Model} \\ \tableline
Index      & 2.00$^{+0.15}_{-0.17}$ & 2.24$^{+0.18}_{-0.12}$ & 2.08$^{+0.12}_{-0.13}$ & & \\ \tableline
$\chi^2_v$ & 1.60 & 0.98 & 1.29 &                 &               \\ \tableline
\multicolumn{6}{|c|}{One-Temperature MEKAL Model} \\ \tableline
$T_1$      & 2.84$^{+0.85}_{-0.61}$ & 2.03$^{+0.52}_{-0.30}$ & 2.25$^{+0.60}_{-0.35}$ & & \\ \tableline
$\chi^2_v$ & 1.82 & 1.47 & 1.59 &                 &               \\ \tableline
\multicolumn{6}{|c|}{Two-Temperature MEKAL Model} \\ \tableline
$T_1$      & 3.43$^{+2.20}_{-0.90}$ & 2.26$^{+0.95}_{-0.50}$ & 3.47$^{+2.50}_{-1.00}$ & & \\ \tableline
$T_2$      & 0.34$^{+0.19}_{-0.09}$ & 0.30$^{+0.21}_{-0.11}$ & 0.38$^{+0.10}_{-0.27}$ & & \\ \tableline
$R_{thermal}$  & 0.29 & 0.35 & 0.375 &    &   \\ \tableline   
$\chi^2_v$ & 1.54 & 1.15 & 1.08 &                 &               \\ \tableline
\end{tabular}
\end{center}
\end{table}

\clearpage

\begin{table}
\begin{center}
\caption{Thermodynamic Properties of the Shell, Lobe, and ISM.}\label{presstab}
\begin{tabular}{|l|c|c|c|}\tableline
Feature & Pressure (dyn cm$^{-2}$) & Temperature (keV) & Density (cm$^{-3}$) \\ \tableline
ISM (region 4)       & 1.0$\times$10$^{-12}$   & 0.29 & 1.7$\times$10$^{-3}$  \\ \tableline
Shell (region 2)     & 2.1$\times$10$^{-10}$   & 2.88 & 2.0$\times$10$^{-2}$  \\ \tableline
Lobe (equipartition) & 1.4$\times$10$^{-11}$   &      &                       \\ \tableline
\end{tabular}
\end{center}
\end{table}

\clearpage


\begin{thebibliography}{}

  \bibitem[Alexander (2002)]{ale02} Alexander, P. 2002, \mnras, {\bf 335}, 610.

  \bibitem[Bahcall \& Sarazin (1977)]{bah77} Bahcall, J. N., and Sarazin, C. L. 1977, \apj, {\bf 213}, L99.

  \bibitem[Blanton \etal(2001)]{bla01r} Blanton, E. L., Sarazin, C. L, McNamara, B. R., and Wise, M. W. 2001, \apj, {\bf 558}, L15.

  \bibitem[Blanton \etal(2001)]{bla01} Blanton, E. L., Sarazin, C. L., and Irwin, J. A. 2001, \apj, {\bf 552}, 106.

  \bibitem[B\"{o}hringer \etal (1993)]{boh93} B\"{o}hringer, H., Voges, W., Fabian, A. C., Edge, A. C., and Neumann, D. M. 1993, \mnras, {bf 264}, L25.

  \bibitem[B\"{o}hringer \etal(2001)]{boh01} B\"{o}hringer, H., \etal 2001, \aap, {\bf 365}, L181.

  \bibitem[B\"{o}hringer \etal(2002)]{boh02} B\"{o}hringer, H., Matsushita, K., Churazov, E., Ikebe, Y, and Arnaud, M. 2001, \aap , {\bf 382}, 804.

  \bibitem[Brighenti \& Mathews (2002)]{bri02} Brighenti, F., and Mathews, W. G. 2002, \apj, {\bf 573}, 542.

  \bibitem[Brown \& Bregman (2001)]{bro01} Brown, B. A. and Bregman, J. N. 2001, \apj, {\bf 547}, 154.

  \bibitem[Br\"{u}ggen \etal (2002)]{brg02m} Br\"{u}ggen, M., Kaiser, C. R., Churazov, E., and En{\ss}lin, T. A. 2002, \mnras, {\bf 331}, 545.

  \bibitem[Br\"{u}ggen \& Kaiser (2002)]{brg02n} Br\"{u}ggen, M., and Kaiser, C. R. 2002, Nature, {bf 418}, 301.

  \bibitem[Brunetti \etal(2002)]{bru02} Brunetti, G., Bondi, M., Comastri, A., and Setti, G. 2002, \aap, {\bf 381}, 795.

  \bibitem[Buote (2001)]{buo01} Buote, D. A. 2001, \apj, {\bf 548}, 642.

  \bibitem[Burns, Feigelson, \& Schreier(1983)]{bur83} Burns, J. O., Feigelson, E. D., and Schreier, E. J. 1983, \apj, {\bf 273}, 128 (BFS83).

  \bibitem[Canizares \etal (1980)]{can87} Canizares, C. R., Fabbiano, G., and Trinchieri, G. 1987, \apj, {\bf 312}, 503.

  \bibitem[Cawley (1998)]{caw98} Cawley, L. 1998, Ph.D. dissertation, The Pennsylvania State University.

  \bibitem[Chevalier (1974)]{che74} Chevalier, R. A. 1974, \apj, {\bf 188}, 501.

  \bibitem[Churazov \etal (1997)]{chu96} Churazov, E., Gilfanov, M., Forman, W., and Jones, C. 1996, \apj, {\bf 471}, 673.

  \bibitem[Ciotti \& Ostriker (1997)]{cio97} Ciotti, L., and Ostriker, J. P. 1997, \apj, {\bf 497}, L105.

  \bibitem[Ciotti \& Ostriker (2001)]{cio01} Ciotti, L., and Ostriker, J. P. 2001, \apj, {\bf 551}, 131.

  \bibitem[Chandra Proposers' Guide (2001)]{pog03} {\em Chandra} Proposers' Observatory Guide, rev. 3.0, 2001.

  \bibitem[Chiaberge, Capetti, \& Celotti (2001)]{chi01} Chiaberge, M., Capetti, A., and Celotti, A. 2001, \mnras, {\bf 324}, L33.

  \bibitem[Churazov \etal (2001)]{chu01} Churazov, E., Br\"{u}ggen, M., Kaiser, C. R., B\"{o}hringer, H., and Forman, W. 2001, \apj, {\bf 554}, 261.

  \bibitem[Clarke, Burns, \& Norman (1992)]{cla92} Clarke, D. A., Burns, J. O., and Norman, M. L. 1992, \apj, {\bf 395}, 444.

  \bibitem[Clarke, Harris, \& Carilli (1997)]{cla97} Clarke, D. A., Harris, D. E., and Carilli, C. L. 1997, \apj, {\bf 284}, 981.

  \bibitem[Cooper, Price, and Cole (1965)]{coo65} Cooper, B. F. C., Price, R. M., and Cole, D. J. 1965, Aust. J. Phys., {\bf 18}, 589.

  \bibitem[Cowie \& McKee (1977)]{cow77} Cowie, L. L., and McKee, C. F. 1977, \apj, {\bf 211}, 135.

  \bibitem[David \etal(1990)]{dav90} David, L. P., Arnaud, K. A., Forman, W., and Jones, C. 1990, \apj, {\bf 356}, 32.

  \bibitem[David \etal(2001)]{dav01} David, L. P., Nelson, P. E. J., McNamara, B. R., Forman, W. R., Jones, C., Ponman, T., Robertson, B. and Wise, M. 2001, \apj, {\bf 557}, 546.

  \bibitem[D\"{o}bereiner \etal(1996)]{dob96} D\"{o}bereiner, S., \etal~1996, \apj, {\bf 470}, L15.

  \bibitem[Dufour \etal(1979)]{duf79} Dufour, R. J., van den Bergh, S., Harvel, C. A., Martins, D. M., Schiffer, F. H. III, Talbot, R. J. Jr., Talent, D. L., and Wells, D. C. 1979, \aj, {\bf 84}, 284.

  \bibitem[Dyson, Falle, \& Perry (1980)]{dys80} Dyson, J. E., Falle, S. A. E. G., and Perry, J. J. 1980, \mnras, {\bf 191}, 785.

  \bibitem[Eskridge, Fabbiano, \& Kim (1995)]{esk95} Eskridge, P. B., Fabbiano, G, and Kim, D. W. 1995, \apjs, {\bf 533}, 799.

  \bibitem[Fabian \etal(2000)]{fab00} Fabian, A. C. \etal~2000, \mnras, {\bf 318}, L65.

  \bibitem[Fabricant, Lecar, \& Gorenstein (1980)]{fab80} Fabricant, D., Lecar, M., and Gorenstein, P. 1980, \apj, {\bf 241}, 552.

  \bibitem[Fabbiano (1989)]{fab89r} Fabbiano, G. 1989, \araa, {\bf 27}, 87.

  \bibitem[Fabbiano, Gioia, \& Trinchieri (1989)]{fab89} Fabbiano, G., Gioia, I. M., and Trinchieri, G. 1989, \apj, {\bf 347}, 127.

  \bibitem[Feigelson \etal(1981)]{fei81} Feigelson, E. D., Schreier, E. J., Delvaille, J. P., Giacconi, R., Grindlay, J. E., and Lightman, A. P. 1981, \apj, {\bf 251}, 31.

  \bibitem[Feigelson \etal(1995)]{fei95} Feigelson, E. D., Laurent-Muehleisen, S. A., Kollgard, R. I., and Fomalont, E. B. 1995, \apj, {\bf 449}, L149.

  \bibitem[Feretti \etal (1999)]{fer99} Feretti, L., Perley, R., Giovannini, G., and Andernach, H. 1999, \aap, {\bf 341}, 29.

  \bibitem[Finoguenov \& Jones (2000)]{fin00} Finoguenov, A. and Jones C. 2000, \apj, {\bf 539}, 603.

  \bibitem[Finoguenov \& Jones (2001)]{fin01} Finoguenov, A. and Jones C. 2001, \apj, {\bf 547}, L107.

  \bibitem[Forman, Jones, \& Tucker (1985)]{for85} Forman, W., Jones, C., and Tucker, W. 1985, \apj, {\bf 293}, 102.

  \bibitem[Fossati \etal(1998)]{fos98} Fossati, G. Maraschi, L., Celotti, A., Comastri, A., and Ghisellini, G. 1998, \mnras, {\bf 299}, 435.

  \bibitem[Garmire \etal(1992)]{gar92} Garmire, G. P., Nousek, J. A., Apparao, K. M. V., Burrows, D. N., Fink, R. L., and Kraft, R. P. 1992, \apj, {\bf 399}, 694.

  \bibitem[Graham (1979)]{gra79} Graham, J. A. 1979, \apj, {\bf 232}, 60.

  \bibitem[Hardcastle, Birkinshaw, \& Worrall (1998)]{har98} Hardcastle, M. J,, Birkinshaw, M., and Worrall, D. M., 1998, \mnras, {\bf 294}, 615.

  \bibitem[Hardcastle, Birkinshaw, \& Worrall (1998)]{har98b} Hardcastle, M. J,, Birkinshaw, M., and Worrall, D. M., 1998, \mnras, {\bf 296}, 1098.

  \bibitem[Hardcastle \& Worrall (2000)]{har00} Hardcastle, M. J., and Worrall, D. M. 2000, \mnras, {\bf 314}, 359.

  \bibitem[Hardcastle, Birkinshaw, \& Worrall (2001)]{har01} Hardcastle, M. J., Birkinshaw, M., and Worrall, D. M. 2001, \mnras, {\bf 326}, 1499.

  \bibitem[Hardcastle \etal (2002)]{hrd02} Hardcastle, M. J., Worrall, D. M., Birkinshaw. M., Laing, R. A., and Bridle, A. H. 2002, \mnras, {\bf 334}, 182.

  \bibitem[Hardcastle \etal (2002)]{hrd02b} Hardcastle, M. J., Birkinshaw, M., Cameron, R., harris, D., Looney, Worrall, D. M. 2002, \apj, in press.

  \bibitem[Harris \etal (2000)]{dhr00} Harris, D. E. \etal~2000, \apj, {\bf 530}, L81.

  \bibitem[Harris \& Krawczynski (2002)]{har02} Harris, D. E. and Krawczynski, H. 2002, \apj, {\bf 565}, 244.

  \bibitem[Heinz, Reynolds, \& Begelman (1998)]{hei98} Heinz, S., Reynolds, C. S., and Begelman, M. C. 1998, \apj, {\bf 501}, 126.

  \bibitem[Hui \etal (1995)]{hui95} Hui, X., Ford, H. C., Freeman, K. C., and Dopita, M. A. 1995, \apj, {\bf 449}, 592. 

  \bibitem[Icke (1992)]{ick92} Icke, V. 1992, 'Beams and Jets in Astrophysics', ed. Philip A. Hughes, Cambridge University Press, Cambridge.

  \bibitem[Israel (1998)]{isr98} Israel, F. P. 1998, Astron. Astrophys. Rev., {\bf 8}, 237.

  \bibitem[Jansen \etal (2001)]{jan01} Jansen, F. \etal~2001, \aap, {\bf 365}, L1.
 
  \bibitem[Jones \etal(1996)]{jon96} Jones, D. L. \etal~1996, \apj, {\bf 466}, L63.
 
  \bibitem[Jones \etal(2002)]{jon02} Jones, C., Forman, W., Vikhlinin, A., Markevitch, M., David, L., Warmflash, A., Murray, S., and Nulsen, P. E. J. 2002, \apj, {\bf 567}, L115.

  \bibitem[Kaiser \& Alexander (1999)]{kai99} Kaiser, C. R., and Alexander, P. 1999, \mnras, {\bf 305}, 707.

  \bibitem[Karovska \etal(2002)]{kar02} Karovska, M., Fabbiano, G., Nicastro, F., Elvis, M., Kraft, R., and Murray, S. 2002, \apj, {\bf 577}, 114.

  \bibitem[Kraft \etal(2000)]{kra00} Kraft, R. P. \etal~2000, \apj, {\bf 531}, L9.

  \bibitem[Kraft \etal(2001)]{kra01} Kraft, R. P., Kregenow, J. M, Jones, C., Forman, W. R., and Murray, S. S. 2001, \apj, {\bf 560}, 675.

  \bibitem[Kraft \etal(2002a)]{kra02} Kraft, R. P., Forman, W. R., Jones, C., Murray, S. S., Hardcastle, M. J., Worrall, D. M. 2001, \apj, {\bf 569}, 54.

  \bibitem[Landau \& Lifschitz (1990)]{lan90} Landau, L. D., and Lifshitz, E. M. 1990, Fluid Mechanics Vol. 6 (Oxford, Butterworth-Heinemann).

  \bibitem[Laing \& Bridle (2002)]{lai02} Laing, R. A., and Bridle, A. H. 2002, preprint, astro-ph 0206251.

  \bibitem[Leahy and Perley (1991)]{lea91} Leahy, J. P. and Perley, R. A. 1991, \aj, {\bf 102}, 537.

  \bibitem[Leahy (1993)]{lea93} Leahy, P. 1993, ``Jets in Extragalactic Radio Sources'', eds. R\"{o}ser, H.-J., and Meisenheimer, K., Berlin: Springer-Verlag.

  \bibitem[Ledlow \& Owen (1996)]{led96} Ledlow, M. J., and Owen, F. N. 1995, \aj, {\bf 109}, 853.

  \bibitem[Loewenstein \& Mushotzky (2002)]{loe02} Loewenstein, M., and Mushotzky, R. F. 2002, BAAS, {\bf 199}, 25.05.

  \bibitem[Mathieu \etal (1996)]{mat96} Mathieu, A., Dejonge, H., and Hui, X. 1996, \aap, {\bf 309}, 30.

  \bibitem[Matsumoto \etal(1997)]{mat97} Matsumoto, H., Koyama, K., Awaki, H., Tsuru, T., Loewenstein, M., and Matsushita, K. 1997, \apj, {\bf 482}, 142.

  \bibitem[McNamara \etal(2000)]{mcn00} McNamara, B. R. \etal~2000, \apj, {\bf 534}, L135.

  \bibitem[Molendi \& Pizzolato (2001)]{mol01} Molendi, S., and Pizzolato, F. 2001, \apj, {\bf 560}, 294.

  \bibitem[Morganti \etal (1999)]{mor99} Morganti, R., Killeen, N. E. B., Ekers, R. D., and Osterloo, T. A. 1999, \aap, {\bf 307}, 750.

  \bibitem[Morini, Anselmo, \& Molteni (1989)]{mor89} Morini, M., Anselmo, F., and Molteni, D. 1989, \apj, {\bf 347}, 750.

  \bibitem[Narayan \& Medvedev (2001)]{nar01} Narayan, R., and Medvedev, M. V. 2001, \apj, {\bf 562}, L129.

  \bibitem[Nulsen \etal (2002)]{nul02} Nulsen, P. E. J., David. L. P., McNamara, B. R., Jones., C., Forman, W. R., and Wise, M. 2002, \apj, {\bf 568}, 163.

  \bibitem[Peterson \etal (2001)]{pet01} Peterson, J. R., Paerels, F. B. S., Kaastra, J. S., Arnaud, M., Reiprich, T. H., Fabian, A. C., Mushotzky, R. F., Jernigan, J. G., Sakelliou, I. 2001, \aap, {\bf 365}, L104.

  \bibitem[Predehl and Schmitt (1995)]{pre95} Predehl, P., and Schmitt, J. H. M. M. 1995, \aa, {\bf 293}, 889.

  \bibitem[Reynolds, Heinz, \& Begelman (2001)]{rey01} Reynolds, C. S, Heinz, S., and Begelman, M. C. 2001, \apj, {\bf 549}, 179.

  \bibitem[Sarazin, Irwin, \& Bregman (2001)]{sar01} Sarazin, C. L., Irwin, J. A., and Bregman, J. N. 2001, \apj, {\bf 556}, 533.

  \bibitem[Saxton, Sutherland, \& Bicknell (2001)]{sax01} Saxton, C. J., Sutherland, R. S., and Bicknell, G. V. 2001, \apj, {\bf 563}, 103.

  \bibitem[Scarpa \& Urry (2001)]{sca01} Scarpa, R., and Urry, C. M. 2001, \apj, {\bf 556}, 749.

  \bibitem[Schiminovich \etal(1994)]{sch94} Schiminovich, D., van Gorkom, J. H.,van der Hulst, J. M., and Kasow, S. 1994, \apj, {\bf 423}, L101.

  \bibitem[Schreier \etal(1996)]{sch96} Schreier, E. J., Capetti, A., Macchetto, F., Sparks, W. B., Ford, H. J. 1996, \apj, {\bf 459}, 535.

  \bibitem[Sedov (1959)]{sed59} Sedov, L. I., 1959, 'Similarity and Dimensional Methods in Mechanics', translated by A. G. Volkovets, CRC Press, 10th ed.

  \bibitem[Slee \etal (1983)]{sle83} Slee, O. B., Sheridan, K. V., Dulk, G. A., Little, A. G. 1983, Proc. ASA, {\bf 5}, 247.

  \bibitem[Skibo, Dermer, \& Kinzer (1994)]{ski94} Skibo, J. G., Dermer, C. D., and Kinzer, R. L. 1994, \apj, {\bf 426}, L23.

  \bibitem[Snowden \etal(1997)]{sno97} Snowden, S. L., Egger, R., Freyberg, M. J., McCammon, D., Plucinsky, P. P., Sanders, W. T., Schmitt, J. H. M. M., Tr\"{u}mper, J., and Voges, W. 1997, \apjs, {\bf 485}, 125.

  \bibitem[Spitzer (1998)]{spi98} Spitzer, L. 1998, Physical Processes in the Interstellar Medium (New York: John Wiley and Sons).

  \bibitem[Stark \etal (1992)]{sta92} Stark, A. A., Gammie, C. F., Wilson, R. W., Bally, J., Linke, R. A., Heiles, C., and Hurwitz, M. 1992, \apjs, {\bf 78}, 77.

  \bibitem[O'Sullivan, Forbes, \& Ponman (2001)]{sul01} O'Sullivan, E., Forbes, D. A., and Ponman, T. J. 2001, \mnras, {\bf 328}, 461.

  \bibitem[Padovani \& Urry (1991)]{pad91} Padovani, P., and Urry, C. M. 1991, \apj, {\bf 368}, 373.

  \bibitem[Tashiro \etal(1998)]{tas98} Tashiro, M., Kaneda, H., Makishima, K., Iyomoto, N., Idesawa, E., Ishisaki, Y., Kotani, T., Takahashi, T., and Yamashita, A. 1998, \apj, {\bf 499}, 713.

  \bibitem[Tingay \etal(1998)]{tin98} Tingay, S. J., \etal~1998, A. J., {\bf 115}, 960.

  \bibitem[Trinchieri \& Fabbiano (1985)]{tri85} Trinchieri, G. and Fabbiano, G., 1985, \apj, {\bf 296}, 447.

  \bibitem[Trinchieri, Fabbiano, \& Canizares (1986)]{tri86} Trinchieri, G., Fabbiano, G., and Canizares, C. R. 1986, \apj, {\bf 310}, 637.

  \bibitem[Trinchieri \& Fabbiano (1991)]{tri91} Trinchieri, G. and Fabbiano, G.~1991, \apj, {\bf 382}, 82.

  \bibitem[Trinchieri \& Di Serego Alighieri (1991)]{tri91b} Trinchieri, G. and Di Serego Alighieri, S. 1991, \aj, {\bf 101}, 1647.

  \bibitem[Trinchieri, Noris, \& Di Serego Alighieri (1997)]{tri97} Trinchieri, G., Noris, L., and Di Serego Alighieri, S. 1997, \aa, {\bf 326}, 565.

  \bibitem[Trinchieri \etal(2000)]{tri00} Trinchieri, G. \etal~2000, \aap, {\bf 364}, 53.

  \bibitem[Tucker \& David (1997)]{tuc97} Tucker, W., and David, L. P. 1997, \apj, {\bf 484}, 602.

  \bibitem[Urry \& Padovani (1995)]{urr95} Urry, C. M. and Padovani P. 1995, \pasp, {\bf 107}, 803.

  \bibitem[van Gorkom \etal (1990)]{gor90} van Gorkom, J. H., van der Hulst, J. M., Haschick, A. D., Tubbs, A. D. 1990, \aj, {\bf 99}, 1781.

  \bibitem[Vikhlinin, Markevitch, \& Murray (2001)]{vik01} Vikhlinin, A., Markevitch, M., and Murray, S. S. 2001, \apj, {\bf 551}, 160.

  \bibitem[Weisskopf \etal(2000)]{wei00} Weisskopf, M. C., Tananbaum, H. D., Van Speybroeck, L. P., and O'Dell, S. L. 2000, Proc. SPIE 'X-ray Optics, Instrument ation and Missions III', eds. J. E. Tr\"{u}mper and B. Aschenbach, 2.

  \bibitem[Worrall, Birkinshaw, \& Hardcastle (2001)]{wor01} Worrall, D. M., Birkinshaw, M., and Hardcastle, M. J. 2001, \mnras, {\bf 326}, L7.

  \bibitem[Xu \etal (2001)]{xu01} Xu, H., Kahn, S. M., Peterson, J. R., Behar, E., Paerels, F. B. S., Mushotzky, R. F., Jernigan, J. G., and Makishima, K. 2001, preprint (astro-ph 0110013).

\end{thebibliography}
\end{document}